\documentstyle[12pt,epsfig,amsfonts]{article}
\def \sech{\mathop{\rm sech}\nolimits}

\newcommand{\beq}{\begin{eqnarray}}
\newcommand{\eeq}{\end{eqnarray}}
\newcommand{\eqn}{\begin{equation}}
\newcommand{\een}{\end{equation}}
\begin{document}
\title {One-loop corrections to classical masses \\
      of \\ kink families}
\author{A. Alonso Izquierdo$^{(a,c)}$, W. Garc\'{\i}a Fuertes$^{(b)}$,
M.A. Gonz\'alez Le\'on$^{(c)}$ \\ and J. Mateos Guilarte$^{(d)}$
\\ {\normalsize {\it $^{(a)}$ DAMPT, Cambridge University, UK}}\\{\normalsize
{\it $^{(b)}$ Departamento de F\'{\i}sica}, {\it Universidad de
Oviedo, SPAIN}}\\ {\normalsize {\it $^{(c)}$ Departamento de
Matem\'atica Aplicada}, {\it Universidad de Salamanca, SPAIN}}\\
{\normalsize {\it $^{(d)}$ Departamento de F\'{\i}sica}, {\it
Universidad de Salamanca, SPAIN}}}

\date{}
\maketitle
\begin{abstract}
One-loop corrections to kink masses in a family of
(1+1)-dimensional field theoretical models with two real scalar
fields are computed. A generalized DHN formula applicable to
potentials with and without reflection is obtained. It is shown
how half-bound states arising in the spectrum of the second order
fluctuation operator for one-component topological kinks and the
vacuum play a central r$\hat{{\rm o}}$le in the computation of the
kink Casimir energy. The issue of whether or not the kink
degeneracy exhibited by this family of models at the classical
level survives one-loop quantum fluctuations is addressed .

\end{abstract}

\section{Introduction}

BPS states arising both in extended supersymmetric gauge theories,
\cite{Olive}, and string/M theory, \cite{Duff}, play a crucial
r\^ole in the understanding of dualities between the different
regimes of these systems. In this framework, domain walls appear
as extended states in N=1 SUSY gluodynamics and the Wess-Zumino
model, \cite{Dvali}. It is desirable to compute the one-loop
quantum corrections, $\Delta M$, to the masses of these new
entities. The huge number of fields involved in these theories,
however, renders the task impossible. In a search for inspiration
about this problem, study of quantum corrections to the masses of
(1+1)-dimensional real solitons has been reignited in recent
years, both in a supersymmetric and a purely bosonic framework.

This topic was first addressed in the classic papers of Dashen,
Hasslacher and Neveu, \cite{Dashen}, and Faddeev and Korepin,
\cite{Fad} . The authors treated the purely bosonic $\lambda
[\phi]^4_2$ and sine-Gordon models for a single real scalar field.
In the first model, the quantum correction to the kink mass was
established to be $\Delta M=\hbar m
\left(\frac{1}{2\sqrt{6}}-\frac{3}{\sqrt{2} \pi} \right)$. The
response given in those papers is currently accepted to be
correct. In the eighties, the supersymmetric extension of these
theories was studied widely in references such as
\cite{Adda,Rou,Uchi,Kaul,Imbi}.  The main concerns were the
quantum correction, $\Delta M$, to the SUSY kink mass, and whether
or not the quantum Bogomolny bound was saturated. By the end of
the past century consensus about both questions was reached,
mainly based on the papers of the Minnesota, Stony Brook-Vienna
and M.I.T. groups: \cite{Shifman}, \cite{Rebhan} and
\cite{Graham}. In \cite{Shifman} using powerful supersymmetric
techniques a new anomaly in the central charge was found,
following a conjecture in the second paper of \cite{Rebhan}. In
\cite{Rebhan} a profound analysis of the bosonic issue was
achieved. These authors carefully distinguished between the
outcomes obtained using two different kinds of cut-off
regularization procedure. They found that only the regularization
method based on a subtle cut-off in the number of fluctuation
modes fits in smoothly with supersymmetry, precisely the method
that leads to the same result as in the computation performed by
DHN/FK for the bosonic fluctuations. We shall refine the mode
number regularization procedure in a way suitable to be applied
also to potentials with reflection, carefully approaching the
problem from the 1D Levinson theorem as developed in Reference
\cite{Barton}. Dirichlet boundary conditions will be imposed to
manage a discrete spectrum; a cut-off in the energy will be
considered, and a finite number of modes near threshold must be
subtracted in such a way that the numbers of modes counted around
the kink and the vacuum are identical; after this, the continuous
limit is taken.

Graham and Jaffe \cite{Graham} also obtained this response, using
techniques based on continuous phase shift methods. Moreover, they
extended the analysis to the trivial sector in order to compute
the one-loop correction to the static energy of one (far
separated) kink-antikink configuration \footnote{The kink-antikink
system was first correctly analyzed by Schonfeld \cite{Kaul}.}.
Here the authors also notice one important point: half-bound
states in the spectrum of the second order fluctuation operator
around the static configuration require a sharper treatment. Weak
fluctuations around the $\lambda(\phi^4)_2$ and sine-Gordon kinks
are governed by Schr$\ddot{{\rm o}}$dinger operators which
involve potentials without reflection. Thus, the kink zero-point
energy receives a contribution from half-bound states that is
exactly cancelled by the subtraction of identical contributions
from half-bound states to the vacuum Casimir energy. In general,
this is no longer so and extreme care is needed in dealing with
the non-pairing of half-bound states to the kink and the vacuum
sectors.

In both \cite{Bordag} and \cite{Aai0}, generalized zeta function
and heat kernel regularization methods have been used to compute
quantum corrections to the mass of SUSY (the first paper) and
bosonic (the second one) one-component kinks. This latter
procedure directly uses information coming from the potential - no
need to unveil the spectrum - of the Sch$\ddot{{\rm o}}$dinger
operator and one skips the subtleties posed by half- bound states.
In fact, the heat kernel high-temperature expansion, see
\cite{Gilkey}, allowed us to express the one-loop correction to
the $\lambda(\phi^4)_2$-kink mass as an asymptotic series with a
relative error of the 0.07 \%.

The main merit of this approach is the breaking of the number of
field components barrier. So far, only kinks with a single
non-null component were susceptible of being treated
semi-classically; the difficulty lies in the study of the spectrum
of $k\times k$ non-diagonal matrices of differential operators. In
Reference \cite{Aai2}, however, the one-loop correction to the
mass of two-component topological kinks in the celebrated MSTB
model -\cite{MSTB} - has been given as an asymptotic series,
starting from the heat kernel expansion.

In order to be complete we also mention that there are other two
interesting regularization methods to calculate corrections to
kink masses. First, see \cite{Goldhaber}, a local mode
regularization - the cutoff reflecting the spacial variation of
the kink - conceptually improves the mode number method and leads
to the correct answer. Second, see \cite{Rebhan1}, the dimensional
regularization procedure is the appropriate method and has been
successfully used when the kink is embedded in a domain wall.

In this paper we shall discuss a model encompassing two real
scalar fields proposed by Bazeia and coworkers, see \cite{B1},
where the next level of complexity in computing quantum
corrections to kink masses arises. The main novelty with respect
to the MSTB model is the existence of classically degenerate kink
families, see \cite{Voloshin} and \cite{Aai3}. The importance of
the model lies in the fact that Shifman and Voloshin
\cite{Voloshin} have shown this system to be the dimensional
reduction of a generalized Wess-Zumino model with two chiral
super-fields. Thus, in this framework the kink solutions turn into
BPS domain walls of a effective supersymmetric theory. An explicit
demonstration of the stability of some of these solutions is
presented in ref. \cite{Lima} using techniques of SUSY quantum
mechanics. More recently, see \cite{Eto}, it has been shown how
to modify the supersymmetric version of this model to make the
system compatible with local supersymmetry. The kinks of the
(1+1)-dimensional model become exact extended solutions of ${\cal
N}=1$ (3+1)-dimensional supergravity with the local
superpotential proposed by Eto-Sakai. Study of the effective
dynamics arising from quantum fluctuations around these kinks is
of great interest because they become exact domain walls
(two-branes) in ${\cal N}=1$ supergravity.

Whether or not the BPS kinks remain degenerate in mass -or the BPS
domain walls in surface tension- after taking into account
one-loop quantum corrections is the main concern of this
investigation. A clue to answering this question is offered in our
paper \cite{Aai1}. The low-energy dynamics of BPS kinks is shown
to be determined by geodesic motion in the kink moduli space for a
special value of the coupling constant. Bohr-Sommerfeld
quantization of this adiabatic evolution amounts to treating the
Laplace-Beltrami operator of the metric inherited from the kink
kinetic energy as the Hamiltonian. The kink moduli space is the
half-plane and the metric is flat: Therefore, the Hamiltonian is
the ordinary Laplacian for an appropriate choice of coordinates
and it seems that the quantum effects do not lift the kink
degeneracy, at least in this quantum adiabatic limit. We shall
show, however, that these expectations are not fulfilled at the
semi-classical level and repulsive forces between separate lumps
arise.

The organization of the paper is as follows: in Section \S 2 we
introduce the model and briefly describe the structure of the
configuration space and the rich variety of kinks. Section \S 3 is
devoted to computing the quantum correction, $\Delta M$, to the
one-component topological kink TK1 by means of a generalized DHN
procedure applicable to potentials with and without reflection.
The contribution of the half-bound states as a function of the
coupling constant is carefully analyzed. In this Section the
one-loop corrections to the mass of the TK1 kinks are also
estimated, using the zeta function regularization method and the
high-temperature expansion. Comparison between the approximate and
exact results for $\Delta M ({\rm TK1})$ serves as an evaluation
of the error. Starting from numerical solutions, Section \S 4
offers a computation of the semi-classical masses for the whole
BPS kink variety at three special values of the coupling constant.
If $\sigma=2.5$ and $\sigma=1.5$ we find a rapid convergence of
the asymptotic series, although violation of the classical
degeneracy is suggested by our results. The anomalous kink
degeneracy is analyzed in connection with the existence of BPS
link kinks. For $\sigma=2$, the system becomes equivalent to two
decoupled $\lambda [\phi]^4_2$ models and the numerical method
provides results in agreement with our previous work in
\cite{Aai0}: in this case the kink degeneracy still holds at the
one-loop level. We also compute in this Section the one-loop
corrections to the mass of the BPS link kinks for the value
$\sigma=2$ of the coupling-constant in order to easily show that
the kink sum rules hold at the semi-classical level. In Section
\S 5 the dependence of the one-loop kink mass formula on the
parameter c is studied from a general point of view, relying once
again on asymptotic methods (the high-temperature expansion). The
model also has non-BPS kinks. Section \S 6 attempts to elucidate
why non-BPS kinks arise in some regimes of the coupling constant.
In particular, study of stability via index theorems shows that
families of this kind of kinks exist for some critical values of
the coupling constant. Finally, an Appendix is offered explaining
the generalized DHN formula, derived using the cut-off in the mode
number regularization procedure but also valid for potentials with
reflection.

\section{The model: classical kink degeneracy}

We shall focus on the (1+1)-dimensional two-component scalar
field model introduced in \cite{B1}. This system arose as the
bosonic sector of a supersymmetric theory with superpotential:
\[
\bar{W}(\vec{\chi})=\lambda \left( \frac{1}{3} \chi_1^3 -a^2
\chi_1 \right)+\frac{1}{2} \mu \chi_1 \chi_2^2 \qquad .
\]
Here
$\vec{\chi}(y^\mu)=\chi_1(y^\mu)\vec{e}_1+\chi_2(y^\mu)\vec{e}_2$
is a two-component real- scalar field; $y^\mu$, $\mu=0,1$ , are
coordinates in the ${\Bbb R}^{1,1}$ space-time, and $\vec{e}_a$ ,
$a=1,2$ , form an orthonormal basis in the the ${\Bbb R}^2$
internal space : $\vec{e}_a \cdot \vec{e}_b=\delta_{ab}$. We also
choose the metric $g^{\mu\nu}={\rm diag}(1,-1)$ in ${\Bbb
R}^{1,1}$  and a system of units where $c=1$ but keep $\hbar$
explicit. The coupling constants thus have the following units: $
[\lambda]=[\mu]= M^{-\frac{1}{2}} L^{-\frac{3}{2}}$ and $[a^2]=
ML$. The dimension of the scalar field is:
$[\vec{\chi}]=M^{\frac{1}{2}}L^{\frac{1}{2}}$ .

The dynamics is governed by the action:
\begin{equation}
\bar{S}[\vec{\chi}]= \int d^2 y \left[ \frac{1}{2}
\partial_\mu\vec{\chi} \cdot
\partial^\mu \vec{\chi}- \frac{1}{2} \vec{\nabla}\bar{W}
\cdot \vec{\nabla}\bar{W} \right]\qquad .
 \label{eq:action}
\end{equation}
Introducing dimensionless fields, variables and parameters
$\vec{\chi} = 2 a \vec{\phi}$, $y^1=\frac{x}{a \lambda}$,
$y^0=\frac{t}{a\lambda}$, and $\sigma=\frac{\mu}{\lambda}$, we
obtain expressions which are simpler to handle.

For static configurations and the dimensionless superpotential
$W(\vec{\phi})=2 \left(\frac{1}{3} \phi_1^3-\frac{1}{4}
\phi_1+\frac{\sigma}{2} \phi_1 \phi_2^2 \right)$ the energy
functional reads:
\begin{eqnarray}
&& \bar{\cal E}[\chi_1,\chi_2]=4 a^3 \lambda \, {\cal
E}[\phi_1,\phi_2]  \nonumber \\ && {\cal E}[\phi]=\int dx \left[
\frac{1}{2} \left(\frac{d \phi_1}{d x}\right)^2 + \frac{1}{2}
\left(\frac{d \phi_2}{d x}\right)^2 + \frac{1}{8} \left( 4
\phi_1^2+ 2 \sigma \phi_2^2-1 \right)^2 + 2 \sigma^2 \phi_1^2
\phi_2^2 \right] \qquad . \label{eq:adimener}
\end{eqnarray}
It is a function of the unique classically relevant coupling
constant $\sigma$ and also depends on the dimensionless potential
term $U(\phi_1,\phi_2)=\frac{1}{8} \left( 4 \phi_1^2+ 2 \sigma
\phi_2^2-1 \right)^2 + 2 \sigma^2 \phi_1^2 \phi_2^2$.

There are four classical vacuum configurations in this system,
organized in two orbits of the internal parity symmetry group
$G_I={\Bbb Z}_2\times {\Bbb Z}_2$ generated by the transformations
$\pi_1:(\phi_1,\phi_2) \rightarrow (-\phi_1,\phi_2)$ and
$\pi_2:(\phi_1,\phi_2) \rightarrow (\phi_1,-\phi_2)$. The vacuum
moduli space, however, includes two points if $\sigma>0$:
\[
{\cal M}=\frac{G_I}{H_I^{(1)}}\cup
\frac{G_I}{H_I^{(2)}}\hspace{0.5cm},\hspace{0.5cm}\bar{\cal
M}=\frac{{\cal M}}{G_I}={\rm point}\cup {\rm point}
\]
\[
\vec{\phi}_{V_1^{\pm}}(x,t)=\pm \frac{1}{2} \vec{e}_1
\hspace{1cm},\hspace{1cm}\vec{\phi}_{V_2^{\pm}}(x,t)=\pm
\frac{1}{\sqrt{2\sigma }} \vec{e}_2 \qquad .
\]
Here, $H_I^{(a)}$ is the sub-group of $G_I$ that leaves the
$\vec{\phi}_{V_a^{\pm}}$ point invariant. If $\sigma<0$, the
$\vec{\phi}_{V_2^{\pm}}$ is lost and the vacuum moduli space has
only one point in this regime.

\noindent\begin{figure}[htbp]
\centerline{\epsfig{file=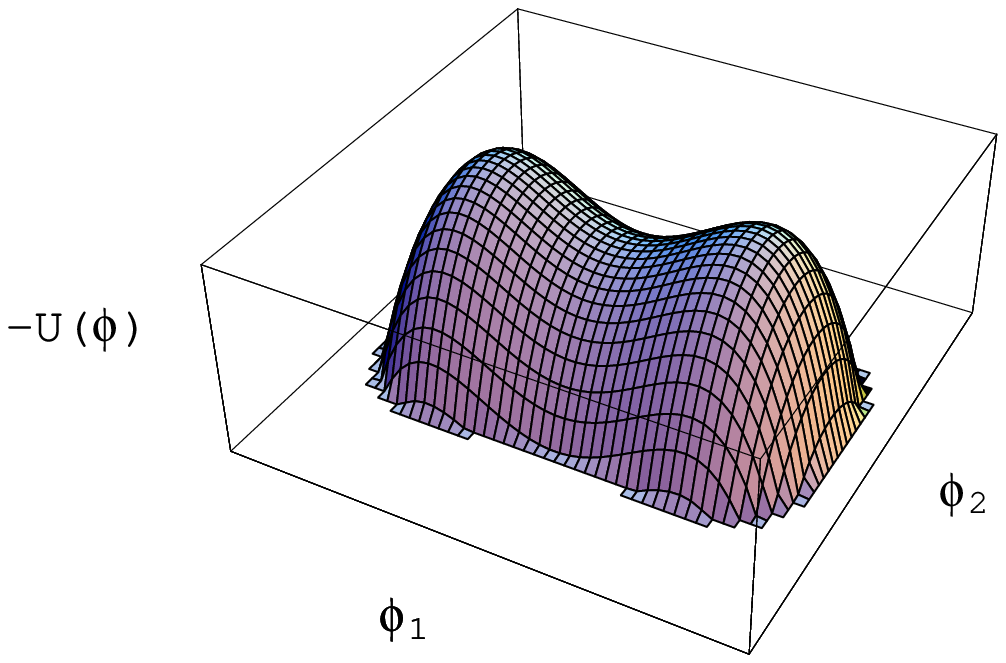, height=4.5cm} \hspace{1.2cm}
\epsfig{file=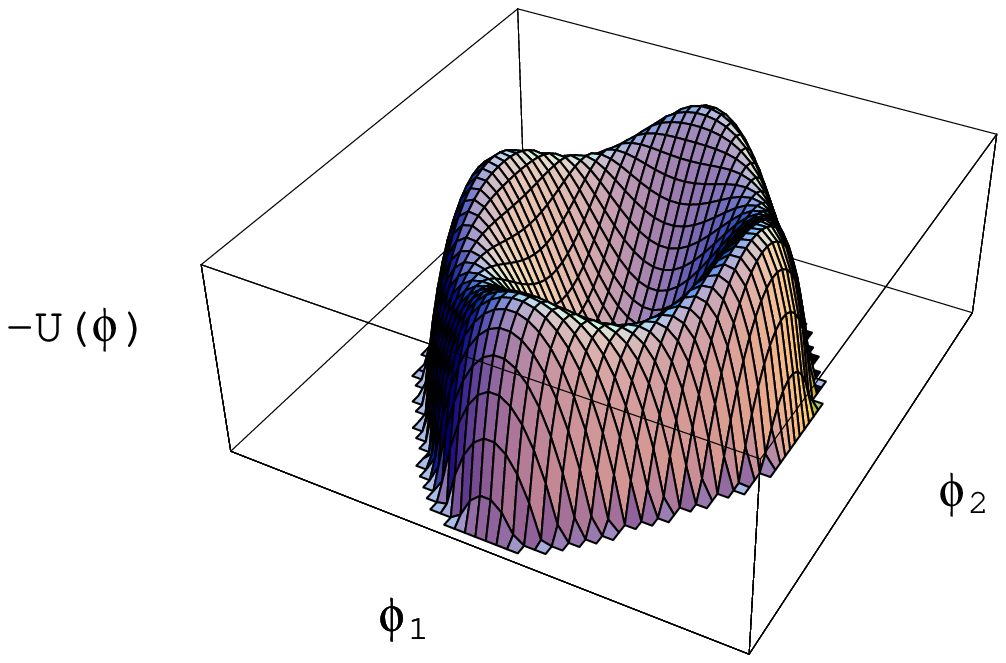,height=4.5cm}} \caption{\small \it The
potential $-U(\phi_1,\phi_2)$ for $\sigma<0$ (left) and for
$\sigma>0$ (right)}
\end{figure}

We shall restrict ourselves here to the $\sigma>0$ regime, because
a richer manifold of kinks arises in this range. The space of
finite energy configurations ${\cal C}$ is the union of sixteen
disconnected topological sectors in this case. If $a,b=1,2$ we
have that:
\[
{\cal C}=\sqcup_{a,b}\quad {\cal C}_{(a,b)}^{\pm\pm}\quad \sqcup_{
a,b}\quad {\cal C}_{(a,b)}^{\pm\mp}
\]
with
\begin{eqnarray*}
 {\cal
C}_{(a,b)}^{\pm\pm}&=& \left\{ \vec{\phi}(x,t)/\vec{\phi}(\infty
,t)=\vec{\phi}_{V_a^\pm},\vec{\phi}(-\infty
,t)=\vec{\phi}_{V_b^\pm}\right\}\\ {\cal
C}_{(a,b)}^{\pm\mp}&=&\left\{ \vec{\phi}(x,t)/\vec{\phi}(\infty
,t)=\vec{\phi}_{V_a^\pm},\vec{\phi}(-\infty
,t)=\vec{\phi}_{V_b^\mp}\right\}
\end{eqnarray*}
In the ${\cal C}_{(a,a)}^{\pm\pm}$ non-topological sectors a
symmetry breaking scenario develops: the original $G_I$ symmetry
of the action (\ref{eq:action}) is broken to the $H_I^{(a)}$
sub-group by the choice of $|\vec{\phi}_{V_a^{\pm}}\rangle $ as
the \lq\lq zero-particle" state in the ${\cal C}_{(a,a)}^{\pm\pm}$
sector.

The small fluctuation solutions
$\vec{\eta}_a(t,x)=\vec{\phi}_a^\pm+\delta\eta(t,x)$ of the field
equations
\[
\left(\frac{\partial^2}{\partial t^2}-\frac{\partial^2}{\partial
x^2}\right)\vec{\phi}=-\vec{\nabla}U
\]
around the vacuum points $V_1^{\pm}$ and $V_2^{\pm}$ can be
respectively expanded in terms of the eigenfunctions of the
Hessian operators:
\[
{\cal V}_1=\left( \begin{array}{cc} -\frac{d^2}{dx^2}+4 & 0 \\ 0 &
-\frac{d^2}{dx^2}+ \sigma^2 \end{array} \right) \hspace{2cm} {\cal
V}_2=\left( \begin{array}{cc} -\frac{d^2}{dx^2}+2 \sigma & 0 \\ 0
& -\frac{d^2}{dx^2}+2 \sigma \end{array} \right)\qquad .
\]
Therefore, the one-particle states have respectively
(dimensionless) masses: $m_1^2(V_1^{\pm})=4
,m_2^2(V_1^{\pm})=\sigma^2$ and
$m_1^2(V_2^{\pm})=m_2^2(V_2^{\pm})=2\sigma$.

The most important feature of this system is that there exists a
degenerate in energy family of BPS kinks in the ${\cal
C}_{(1,1)}^{\pm\mp}$ topological sectors for any value of
$\sigma>0$. BPS kinks are finite energy solutions of the
first-order system of differential equations:
\begin{equation}
\frac{d \phi_1}{dx}=\pm\frac{\partial W}{\partial \phi_1}=\pm
\left( 2 \phi_1^2+ \sigma \phi_2^2-\frac{1}{2}\right) \hspace{1cm}
\frac{d \phi_2}{dx}=\pm\frac{\partial W}{\partial \phi_2}=\pm 2
\sigma \phi_1 \phi_2 \qquad . \label{eq:bps}
\end{equation}
In both References \cite{Voloshin} and \cite{Aai3} it has been
shown that the BPS kinks belonging to the ${\cal
C}_{(1,1)}^{\pm\mp}$ topological sectors are in one-to-one
correspondence with the curves in the ${\Bbb R}^2$ internal space:
\begin{itemize}
\item $\sigma \neq 1$,
\begin{equation}
\phi_1^2 +\frac{\sigma}{2(1-\sigma)}
\phi_2^2=\frac{1}{4}+\frac{c}{2 \sigma} |\phi_2|^\frac{2}{\sigma}
\quad , \label{eq:tra1}
\end{equation}
where $c \in (-\infty,c^S)$ is an integration constant and there
is a critical  value  $c^S=\frac{1}{4} \frac{\sigma}{1-\sigma} (2
\sigma)^\frac{\sigma+1}{\sigma}$.
\item  $\sigma = 1$ ,
\[
\phi_1^2-\phi_2^2 \left( \frac{c}{2}+\log |\phi_2|
\right)=\frac{1}{4}\quad ,
\]
$c \in (-\infty,c^S)$ , $c^S=-1+\ln 2$.
\end{itemize}
One sees in Figure 1 (right) that there are two maxima of
$-U(\phi_1,\phi_2)$ with the same height. Kink solutions which go
from one maximum to the other depend on a parameter $c$ which
measures whether the particle moves through the bottom of the
valley, or more along the sides on the curve (\ref{eq:tra1}).
There is a critical value of $c$ where the particle moves as high
as possible; increasing $c$ beyond this critical value the
particle crosses the mountain and falls off to the other side.

All the BPS kinks belonging to these families have the same
(dimensionless) energy: ${\cal E}({\rm
BPS})=|W(\vec{\phi}_{V_1^\pm})-W(\vec{\phi}_{V_1^\mp})|=\frac{1}{3}$.
We shall denote these kinks as TK2(c) because from (\ref{eq:tra1})
one sees that the two components of the field are non-zero for
generic values of $c$; also, all the kink orbits in
(\ref{eq:tra1}) start and end in vacua belonging to the same
$G_I$-orbit, and are accordingly classified as \lq\lq loop" kinks.

Plugging (\ref{eq:tra1}) into (\ref{eq:bps}) one reduces the
solution of the (\ref{eq:bps}) ODE system to a single quadrature.
The explicit integration can be performed analytically only for
certain special values of $\sigma$, see References
\cite{Voloshin}, \cite{Aai3}.

Exactly for $c=-\infty$, one-component topological kinks -
therefore termed as TK1- arise. If $c=-\infty$, for any $\sigma\in
[0,\infty)$, we obtain the curve $\phi_2=0$ in (\ref{eq:tra1}) and
find the BPS kinks in the ${\cal C}_{(1,1)}^{\pm\mp}$ sectors of
configuration space:
\[
\vec{\phi}_{TK1}(x)= (-1)^\alpha \frac{1}{2} \tanh (x+a)\vec{e}_1
\qquad \alpha\in {\Bbb Z}/{\Bbb Z}_2 \qquad .
\]
$a\in {\Bbb R}$ is another integration constant that fixes the
center of the kink.

\section{One-loop correction to the mass of one-component topological kinks}

In this Section we shall compute the one-loop quantum correction
to the classical mass of TK1 kinks. Because the matrix
differential operator governing second-order fluctuations around
these kinks is diagonal,
\begin{equation} {\cal K}=\left( \begin{array}{cc}
-\frac{d^2}{dx^2}+4 -6 \sech^2 x & 0 \\ 0 &
-\frac{d^2}{dx^2}+\sigma^2 -\sigma (\sigma+1)\sech^2 x
\end{array} \right)  \label{eq:hesstk1}
\end{equation}
this task can be performed using both the DHN formula and the
asymptotic method. Note also that the one-loop correction to the
TK1 kink mass is the sum of two contributions: (1) $\Delta M({\cal
K}_{11})$ is the one-loop correction due to the the tangent
fluctuations to the TK1 kink governed by  ${\cal K}_{11}$. (2)
$\Delta M({\cal K}_{22 })$ collects the contributions to the
quantum corrections coming from the orthogonal fluctuations to the
TK1 ruled by ${\cal K}_{22}$. Therefore, we write
\[
\Delta M({\rm TK1})=\Delta M({\cal K}_{11})+\Delta M({\cal
K}_{22}) \qquad .
\]

\subsection{One-loop correction to the TK1 kink mass: DHN formula}

In (\ref{eq:hesstk1}), both ${\cal K}_{11}$ and  ${\cal K}_{22}$
are particular cases of Schr$\ddot{{\rm o}}$dinger operators of
P\"osch-Teller type
\[
\Delta=-\frac{d^2}{dx^2}+c_0^2-\frac{u_0}{{\rm cosh}^2x}\qquad ,
\]
whose eigenvalues and eigenfunctions are known \cite{Drazin}.
Since ${\cal K}_{22}$ is a Schr\"odinger operator involving
potential terms with reflection $(\sigma \notin {\Bbb N})$ and
without reflection $(\sigma \in {\Bbb N})$, we use the
generalized DHN formula for the $\hbar$-correction to the mass of
one-component topological kinks:
\begin{eqnarray}
\Delta M({\cal K}_{aa})&=&\frac{\hbar m}{2} \left[
\sum_{i=0}^{l-1} \omega_i+s_{l}\omega_{l}-\frac{v_a}{2}  +
\frac{1}{\pi} \int_0^\infty \!\!\! dq \frac{\partial
\delta_a(q)}{\partial q} \sqrt{q^2+v_a^2}- \frac{\left< V_{aa}(x)
\right> }{2 \pi} \right]+ \nonumber \\&+& \hbar m\frac{\left<
V_{aa}(x) \right> }{8 \pi} \int_0^\infty \!\!
\frac{dk}{\sqrt{k^2+v_a^2}}\qquad ,\label{eq:dhn}
\end{eqnarray}
where
\[
\displaystyle v_a^2=\left. \frac{\delta^2
U}{\delta\phi_a^2}\right|_{\vec{\phi}_{V^\pm_1}}
\hspace{0.5cm};\hspace{0.5cm} V_{aa}(x) =
v_a^2-\left.\frac{\delta^2
U}{\delta\phi_a^2}\right|_{\vec{\phi}_{TK1}}\qquad ,
\]
and $\left< \cdot \right>$ stands for the expectation value:
$\left<V_{aa}(x)\right>=\int_{-\infty}^\infty \!dx V_{aa}(x)$.

Formula (\ref{eq:dhn}) has been derived in the Appendix collecting
previous work in this topic. We set as $m=a\lambda$ the parameter
with dimension of $L^{-1}$ that will be used to fix the real
dimension of each observable in the system. Note that as a
consequence of our choice of dimensionless variables there is a
global factor of $\sqrt{2}$ in formula (\ref{eq:dhn}) with respect
to the analogous formula used in \cite{Aai0}. We now pause to
explain the origin of the different terms:
\begin{enumerate}
\item The first line accounts for the zero-point energy of the
quantum kink measured with respect to the zero-point energy of the
vacuum. Eigenstates from both the discrete and continuous spectrum
of ${\cal K}_{aa}$ contribute. The highest bound state contributes
with a weight $s_{l}$, which is 1 if $\omega_{l}$ does not
coincide with the bottom of the continuous spectrum (threshold).
If $\omega_{l}$ is buried at threshold, the corresponding
eigenstate is a half-bound state and $s_{l}=\frac{1}{2}$. To
deduce formula (\ref{eq:dhn}) the density of states in the
continuous spectrum is given in terms of the phase shifts
$\delta_a(q)$ of the scattering waves through the potential:
$v_a^2-V_{aa}(x)$. A cut-off in the number of modes (see Appendix
for a complete derivation of formula (\ref{eq:dhn}) including the
case of potentials with non-zero reflection coefficients) has been
used for renormalizing the zero point energy. Note also that the
subtraction of $\frac{v_a}{2}$ amounts to taking into account the
contribution of the half-bound state in the vacuum to the vacuum
zero point energy (see \cite{Graham} and the Appendix).
\item The second line takes into account the contribution of
the mass renormalization counter-term to the Casimir energies of
both the kink and the vacuum.
\end{enumerate}

\subsubsection{Tangent fluctuations:
$\Delta M({\cal K}_{11})$}

Spec$({\cal K}_{11})$ includes two bound states, one half-bound
state (in the above notation $l=2$), and scattering eigenfunctions
characterized by the phase shifts:
\[
\delta_1 (q)=-2{\rm arctan}\frac{3q}{2-q^2}
\]
Therefore:
\[
{\rm Spec}({\cal K}_{11})=\{\omega_0=0\}\cup
\{\omega_1=3\}\cup\{\omega_2=4\}_{s_2=\frac{1}{2}}\cup
\{q^2+4\}_{q\in {\Bbb R}}\quad .
\]

Substituting all this information into formula (\ref{eq:dhn}) we
obtain: {\small\begin{eqnarray*} \Delta M({\cal K}_{11}^{\rm TK1})
&=& \frac{\sqrt{3}\hbar m}{2}-\frac{\hbar m}{2 \pi}
\int_{-\infty}^\infty \hspace{-0.3cm} dq \frac{3 \sqrt{q^2+4}
(q^2+2)}{q^4+5 q^2+4}+ \frac{3 \hbar m}{2 \pi}
\int_{-\infty}^\infty \frac{d k}{\sqrt{k^2+4}} - \frac{\hbar m}{4
\pi} \int_{-\infty}^\infty \hspace{-0.3cm}dx 6 \sech^2 x
\\&=&\hbar m\left(\frac{1}{2\sqrt{3}}-\frac{3}{\pi} \right)
\end{eqnarray*}}
Except for a global factor $\sqrt{2}$, the contribution of the
tangent fluctuations is the same as the contribution of all the
fluctuations to the kink in the $\phi^4$ model; note that there
are two bound states with eigenvalues of 0 and 3, giving tangent
fluctuations to the ${\rm TK}1$ kink identical to those arising in
the $\lambda(\phi^4)_2$ model. The global factor is due to the
different choice of the parameters in the potential.

\subsubsection{Orthogonal fluctuations: $\Delta M({\cal K}_{22})$}

The scattering in the potential well - $c_0^2=\sigma^2$,
$u_0=\sigma(\sigma+1)$ - of ${\cal K}_{22}$ is not \lq\lq
reflectionless" if $\sigma$ is not an integer. There are in
general even and odd phase shifts
\[
\delta_2^{\pm} (q)=\frac{1}{4}{\rm arctan}\left(\frac{{\rm
Im}(T(q)\pm R(q))}{{\rm Re}(T(q)\pm R(q))}\right) ,
\]
to be read from the transmission and reflection coefficients
\[
T(q)=\frac{\Gamma(\sigma+1-iq)\Gamma(-\sigma-iq)}{\Gamma(1-iq)\Gamma(-iq)}\quad
; \quad
R(q)=\frac{\Gamma(\sigma+1-iq)\Gamma(-\sigma-iq)\Gamma(iq)}
{\Gamma(1+\sigma)\Gamma(-\sigma)\Gamma(-iq)}\qquad .
\]
Recalling that
\begin{equation}
e^{i 2 \delta_2^\pm}=T(q)\pm R(q) \label{eq:hhh}
\end{equation}
$T(q)$ and $R(q)$ are obtained from the asymptotic behaviour of
the scattering eigenstates:
\[
\psi_q(x) = N e^{i q x} {}_2F_1[\textstyle -\sigma,\sigma+1,1-i
q,\frac{e^{x}}{e^{x}+e^{-x}}]\qquad ,
\]
where ${}_2F_1[a,b,c;d]$ is the Gauss hypergeometric function .

If $\sigma=l\in {\Bbb N}$ is a natural number, $R(q)=0$ -
$\Gamma(-\sigma)$ has a pole in this case-,
$\delta_2^+(q)=\delta_2^-(q)$, and the total phase shift
$\delta_2(q)=\delta_2^+(q)+\delta_2^-(q)$ is:
\begin{equation}
\delta_2(q)=\frac{1}{2}{\rm arctan}\left(\frac{{\rm
Im}\prod_{n=0}^{l-1}(q^2-(l-n)^2+2iq(l-n))}{{\rm
Re}\prod_{n=0}^{l-1}(q^2-(l-n)^2+2iq(l-n))}\right). \label{phases}
\end{equation}
The number of bound and half-bound states of Spec$({\cal K}_{22})$
is equal to $I[\sigma]$, $I[\sigma]$ being the integer part of
$\sigma$. The corresponding eigenvalues are
$\omega_i=i(2\sigma-i)$, see \cite{Morse,Drazin}, whereas the
bound state eigenfunctions read:
\[
\psi_i(x) = \frac{N}{({\rm cosh}\,x)^{(\sigma-i) }}
{}_2F_1[\textstyle -i,2\sigma-i+1,\sigma -i+1,\frac{1}{2}(1+{\rm
tanh}x)]\qquad , i=0,1,\cdots  I[\sigma ] \qquad .
\]
The half-bound state arises only in the case $\sigma=I[\sigma]=l
\in {\Bbb N}$. Some explanation about this issue should be
provided: The \lq\lq first" eigenfunction in the continuous
spectrum
\[
\psi_{q=0}(x)=N {}_2F_1[-\sigma,\sigma +1,1; \frac{1}{2}(1+{\rm
tanh}x)]
\]
does not belong to the Hilbert space because
$\lim_{x\rightarrow\pm\infty}\psi_{q=0}(x)\cong c_{\pm}x$, except
if $u_0=l(l+1)$. In this latter case, $\psi_{i=l}=\psi_{q=0}$ ,
\[
\lim_{x\rightarrow\pm\infty}{}_2F_1[{\textstyle -l},{\textstyle
l+1},1;{\textstyle\frac{1}{2}(1+{\rm tanh}x)}]\cong c_{\pm}
\]
and hence this is the half-bound state in the spectrum of ${\cal
K}_{22}$.

In sum, we have that:
\[
{\rm Spec}\, ({\cal K}_{22})= \left\{
\begin{array}{lcl} \cup_{i=0,1,...,I[\sigma]}\{\omega_i= i(2\sigma-i) \} \cup
\{q^2+\sigma^2\}_{q\in {\Bbb R}^+}  && \mbox{if} \hspace{0.4cm}
\sigma \notin {\Bbb N} \\ \cup_{i=0,1,...,\sigma-1}\{\omega_i=
i(2\sigma-i) \} \cup \{\omega_{l=\sigma}=\sigma^2
\}_{s_{l=\sigma}=\frac{1}{2}} \cup \{q^2+\sigma^2\}_{q\in {\Bbb
R}^+}  && \mbox{if} \hspace{0.4cm} \sigma \in {\Bbb N}
\end{array} \right. \qquad .
\]
Therefore, in the subspace of ${\cal C}_{(1, 1)}^{\pm\mp}$
orthogonal to the ${\rm TK1}$ configuration more bound states,
depending in number on the value of $\sigma^2$, appear as
eigen-fluctuations of the Hessian operator. The corresponding
eigenvalues are semi-definite positive for any value of sigma. The
${\rm TK}1$ kink is thus stable independently of the $\sigma^2$
parameter. An important point is that there exists a second zero
mode because the first eigenvalue of ${\cal K}_{22}$ is zero for
all $\sigma^2$. This zero mode obeys neutral equilibrium
fluctuations in a continuous family of kinks with the same energy
as the ${\rm TK}1$ kink and second non-zero component, see
\cite{Aai3}.

Formula (\ref{eq:dhn}) demands the computation of the derivative
of the phase shift with respect to the momentum $q$, using
(\ref{eq:hhh}) can be performed explictly:
\begin{eqnarray}
\frac{\partial \delta_2(q)}{\partial q}&=& -\frac{i}{2}\left[
e^{-i2\delta_2^+} \frac{\partial e^{i 2 \delta_2^+}}{\partial q}+
e^{-i 2 \delta_2^-} \frac{\partial e^{i 2 \delta_2^-}}{\partial
q}\right]= \nonumber \\ &=& 2 Re[\psi(iq)-\psi(-\sigma+i
q)]+\frac{\pi}{2 \sinh^2 \pi q \csc 2 \pi \sigma+\tan \pi
\sigma}\qquad .
 \label{eq:derfase}
\end{eqnarray}
Plugging all these expressions into (\ref{eq:dhn}) we have that:

{\small\begin{eqnarray} && \Delta M({\cal K}_{22})= \nonumber\\
 &&= \frac{\hbar m}{2} \left[ \sum_{i=0}^{I[\sigma]}
\sqrt{i(2\sigma-i)} \!-\!\frac{\sigma}{2}\! +\! \frac{1}{\pi}
\int_0^\infty \hspace{-0.4cm} dq \left( \frac{\partial
\delta_2(q)}{\partial q} \sqrt{q^2\!+\!\sigma^2}
+\frac{\sigma(1+\sigma)}{\sqrt{q^2\!+\!\sigma^2}} \right) \!-\!
\frac{\sigma(\sigma+1) }{\pi} \right] \hspace{0.3cm} \mbox{if}
\hspace{0.3cm} \sigma \notin {\Bbb N} \nonumber \\ &&= \frac{\hbar
m}{2} \left[ \sum_{i=0}^{\sigma-1} \sqrt{i(2\sigma-i)} +\!
\frac{1}{\pi} \int_0^\infty \hspace{-0.4cm} dq \left(
\frac{\partial \delta_2(q)}{\partial q} \sqrt{q^2\!+\!\sigma^2}
+\frac{\sigma(1+\sigma)}{\sqrt{q^2\!+\!\sigma^2}} \right) \!-\!
\frac{\sigma(\sigma+1) }{\pi} \right] \hspace{0.3cm} \mbox{if}
\hspace{0.3cm} \sigma \in {\Bbb N} \qquad . \label{eq:dhn2}
\end{eqnarray}}
We are now interested in describing separately the behaviour of
the contributions from the discrete spectrum of the operator
${\cal K}_{22}$ and from the integral in (\ref{eq:dhn2}) to
$\Delta M({\cal K}_{22})$. We define
\begin{equation}
\Delta M({\cal
K}_{22}^{(d)})=\sum_{i=0}^{I[\sigma]}\sqrt{i(2\sigma-i)}\qquad ,
\qquad f(q,\sigma )=\left( \frac{\partial \delta_2(q)}{\partial q}
\sqrt{q^2\!+\!\sigma^2}
+\frac{\sigma(1+\sigma)}{\sqrt{q^2\!+\!\sigma^2}}
\right)\label{eq:cont}
\end{equation}
to emphasize the different contributions.

In $\Delta M({\cal K}_{22}^{(d)})$, the contribution of the
discrete spectrum of ${\cal K}_{22}$ to the one-loop correction of
the TK1 kink mass, we notice the following. First, the greater the
value of $\sigma$, the greater the number of bound states in ${\rm
Spec}({\cal K}_{22})$ is. There is a jump in $\Delta M({\cal
K}_{22}^{(d)})$ when $\sigma$ crosses an integer value, see Figure
2. Second, a half-bound state in ${\rm Spec}({\cal K}_{22})$
arises exactly at the values $\sigma=l \in {\Bbb N}$, which
contributes to $\Delta M({\cal K}_{22}^{(d)})$ with a
$\frac{1}{2}$ weight with respect to the contribution of any other
state. One would think from these considerations that $\Delta
M({\cal K}_{22})$ is a non-continuous function of the coupling
constant $\sigma$ because of the jumps in $\Delta M({\cal
K}_{22}^{(d)})$. On the other hand, we know that the TK1 solution
depends smoothly on the value of $\sigma$, such that a smooth
response of $\Delta M({\cal K}_{22})$ would be expected.

\noindent\begin{figure}[htbp] \centerline{\epsfig{file=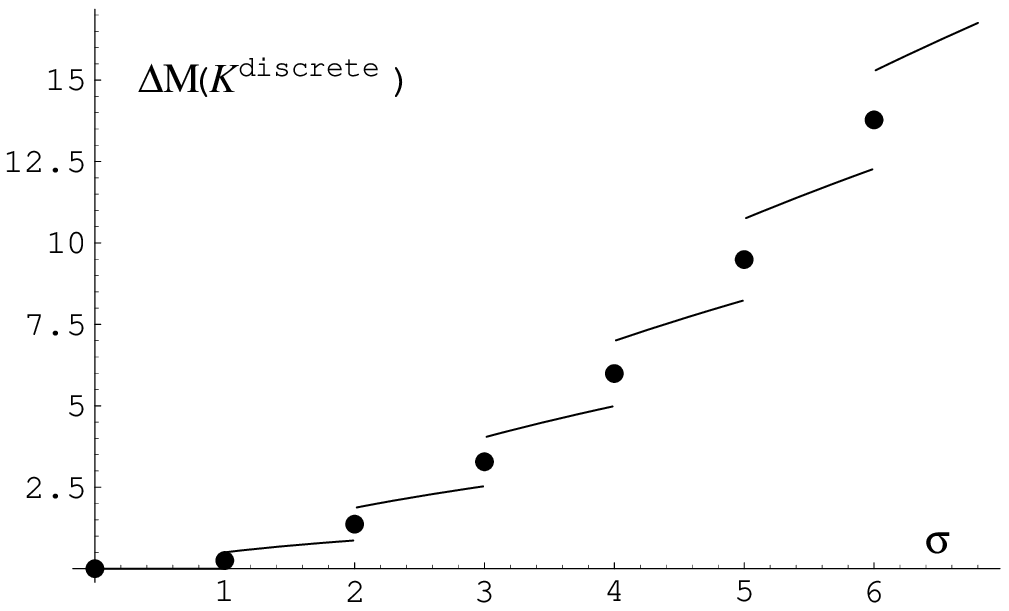,
height=4.cm}\hspace{1cm}\epsfig{file=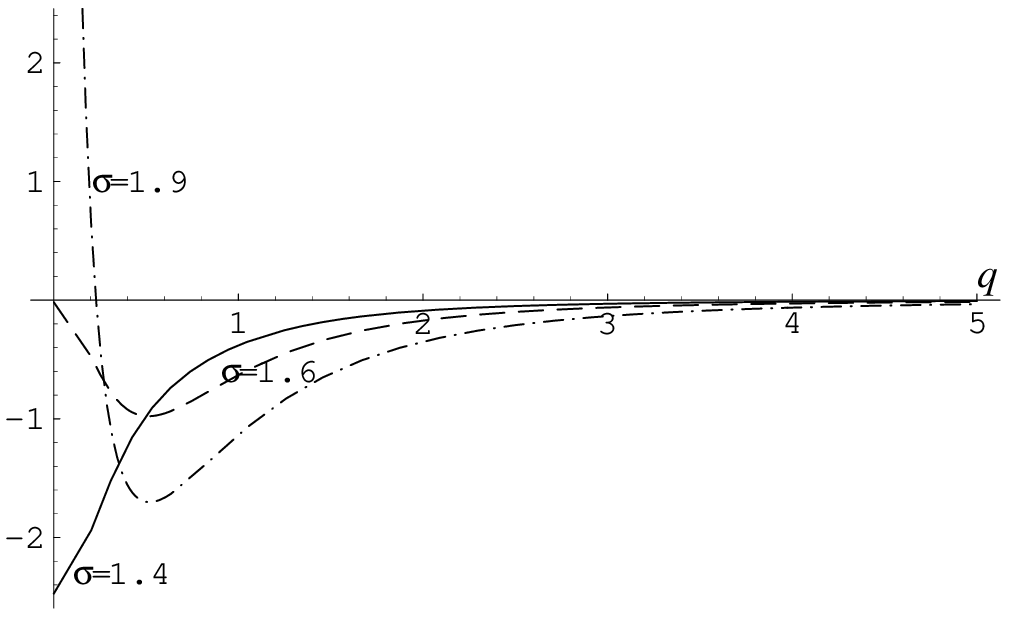, height=4.cm}}
\caption{\small \it Contribution to the one-loop correction from
the discrete spectrum of ${\cal K}_{22}$ (left) Behaviour of the
integrand $f(q,\sigma)$ in (\ref{eq:dhn2}) (right)}
\end{figure}

The analysis of the second contribution is also subtle. The main
points are as follows: First, the descriptive behaviour of the
integrand $f(q,\sigma)$ depends only on the fractional part of
$\sigma$, following the pattern shown in Figure 2 for the range
$\sigma\in (1,2)$. Second, asymptotically we have that:
\[
\lim_{q\rightarrow\infty}f(q,\sigma )=0 \qquad , \quad \forall
\sigma \quad.
\]
Moreover,
\[
\lim_{q\rightarrow 0} f(q,\sigma )=1+\sigma(1-2
\gamma_E-\psi(-\sigma)\psi(1+\sigma))\qquad ,
\]
where $\gamma_E$ is the Euler Gamma constant, and $\psi(z)$ is the
Digamma function. Note that $f(q,\sigma=l)$ has a pole at $q=0$.
Because $f(q,\sigma )$ is well-behaved for any other value of $q$
if $\sigma$ is non-integer, the integral in (\ref{eq:dhn}) is well
defined and a finite answer for $\Delta M({\cal K}_{22})$ is
obtained, see Figure 2. The integral becomes improper if
$\sigma=l\in {\Bbb N}$: in this case it is only defined by taking
the limit $q\rightarrow 0$ appropriately. A miracle happens: the
integral of the function $f(q,\sigma)$ produces a jump in this
contribution when $\sigma $ crosses an integer that exactly
cancels the jump in $\Delta M({\cal K}_{22}^{(d)})$ due to the
appearance of a new bound or half-bound state!, $\Delta M({\cal
K}_{22})$ is a continuous function of $\sigma $.

Nevertheless, the integral cannot be evaluated analytically except
for integer values of $\sigma$. We use numerical analysis to
compute $\Delta M({\cal K}_{22})$ by means of formula
(\ref{eq:dhn}). The numerical computation confirms that $\Delta
M({\cal K}_{22})$ is smooth in $\sigma$. The contributions from
the discrete and continuous spectrum balance each other to obtain
a smooth behaviour in $\sigma$. We stress that this balance is
made possible by a proper counting of the half-bound states in
both the vacuum and TK1 kink Schrodinger operators.

In the next Table we offer the data on the sum of the tangent and
orthogonal contributions to the TK1 quantum mass correction,
$\Delta M({\rm TK1})$, for some values in the range $\sigma \in
(0.4,3.3)$, which are also depicted in Figure 3.

{\footnotesize\begin{center}
\begin{tabular}{|c|c|} \\[-0.6cm] \hline
$\sigma$ & $\Delta M({\rm TK1})/\hbar m$ \\ \hline $0.4$ &
$-0.799335$ \\ $0.5$ & $ -0.829892$ \\ $0.6$ & $-0.860369$ \\
$0.7$ & $ -0.890955$ \\ $0.8$ & $-0.921788$ \\ $0.9$ & $
-0.952966$ \\ $0.99$ & $ -0.981384$ \\ $1.00$ & $ -0.984565$ \\
$1.01$ & $ -0.98775$ \\ $1.1$ & $ -1.01664$ \\ $1.2$ & $ -1.04925$
\\ $1.3$ & $ -1.08242 $ \\ \hline
\end{tabular} \hspace{0.1cm}
\begin{tabular}{|c|c|} \\[-0.6cm] \hline
$\sigma$ & $\Delta M({\rm TK1})/\hbar m$ \\ \hline $1.4$ & $
-1.11618 $ \\ $1.5$ & $ -1.15057 $ \\ $1.6$ & $ -1.18559 $ \\
$1.7$ & $ -1.22128 $ \\ $1.8$ & $ -1.25765 $ \\ $1.9$ & $ -1.2947
$ \\ $1.99$ & $ -1.32865 $ \\ $2.0$ & $-1.33251$  \\ $2.01$ & $
-1.33627 $ \\ $2.1$ &$ -1.37094 $\\ $2.2$ & $ -1.41013 $ \\ $2.3$
& $-1.45005 $ \\ \hline
\end{tabular} \hspace{0.1cm}
\begin{tabular}{|c|c|} \\[-0.6cm] \hline
$\sigma$ & $\Delta M({\rm TK1})/\hbar m$ \\ \hline $2.4$ &
$-1.4907 $ \\ $2.5$ & $ -1.53212$ \\ $2.6$ & $-1.57427 $ \\ $2.7$
& $ -1.61717 $ \\ $2.8$ & $-1.65316 $ \\ $2.9$ & $ -1.70527 $ \\
$2.99$ & $-1.74592 $ \\ $3.0$ & $ -1.72309 $\\ $3.01$ & $ -1.75503
$  \\ $3.1$ & $-1.79644 $  \\ $3.2$ & $ -1.84319 $  \\ $3.3$ & $
-1.89071 $  \\ \hline
\end{tabular}
\end{center}}

\subsection{Asymptotic series for the TK1 semi-classical masses}

 The evaluation of one-loop corrections
 to the masses of other (two-component) kinks which are classically
 degenerate in energy with the TK1 kink is another task to be addressed
 in this work. For this purpose the
 DHN formula is of no use, because our knowledge about the spectrum
 of non-diagonal matrix Sch$\ddot{\rm o}$dinger operators is
 grossly insufficient. Alternatively, we can use a formula, derived in
 References \cite{Aai0,Aai2}, that expresses the one-loop
 correction to $k$-component kink masses as an asymptotic series.
 If we denote as ${\cal K}$ the Hessian operator around a given loop kink
 K, the formula for the one-loop correction to the masses of
topological kinks
 derived in Reference \cite{Aai2} reads:
\begin{equation}
\Delta M ({\cal K}) =\hbar m[\Delta_0 +D_{n_0}] \left\{
\begin{array}{l} \Delta_0=-\displaystyle\frac{j}{2\sqrt{\pi}}  \\
D_{n_0}=-\displaystyle\sum_{a=1}^2\sum_{n=2}^{n_0-1}
\displaystyle\frac{[a_{n}]_{aa}({\cal K})}{8\pi}
\displaystyle\frac{\gamma[n-1,v_a^2]}{v_a^{2n-2}}\quad , \quad
n_0\in{\Bbb N}\qquad .
\end{array}
\right. \label{eq:as}
\end{equation}
A lightning summary of the content of formula (\ref{eq:as}) is as
follows: $j=\mbox{dim Ker}({\cal K})$ is the number of zero modes
in the spectrum of ${\cal K}$. $[a_{n}]_{aa}({\cal K})$ are the
coefficients of the high- temperature expansion of the heat
function associated with the heat equation:
\[
\sum_{b=1}^2 \left(
\frac{\partial}{\partial\beta}\delta_{ab}+{\cal
K}_{ab}\right)F_b(\beta , x) =0 \qquad ,
\]
$\gamma[n-1,v_a^2]$ are incomplete Gamma functions, see
\cite{Abramowitz}, and $v_a$ are the masses of the fundamental
quanta at the point of the vacuum moduli space determined by the
loop kink K. The divergent terms
\[
\sum_{a=1}^2\frac{[a_0]_{aa}({\cal
K})}{8\pi}\frac{\gamma[-1,v_a^2]}{v_a^{-2}}\qquad , \qquad
\sum_{a=1}^2\frac{[a_1]_{aa}({\cal K})}{8\pi}\gamma[0,v_a^2]
\]
lacking in (\ref{eq:as}) are respectively cancelled by subtracting
the zero point vacuum energy, and taking into account the mass
renormalization counterterm, see \cite{Aai0} and \cite{Aai2}.

 Although we have computed $\Delta M (TK1)$ by means of the DHN formula
 (\ref{eq:dhn}) in the previous Section, we
 now apply (\ref{eq:as}) to the TK1 kink in order to test how
 good the approximation provided by the asymptotic series is with
 respect to the exact DHN result.

 The spectrum of ${\cal K}$ comprises two zero modes; henceforth, $j=2$.
Moreover, $v^2_1=4$ and $v_2^2=\sigma^2$ in this case, whereas the
Seeley coefficients $[a_n({\cal K})]_{11}$ and $[a_n({\cal
K})]_{22}$ are pointed out in References \cite{Aai0,Aai2}. The
calculations of the partial sums in (\ref{eq:as}) for several
values of $\sigma $ and with $n_0=10$ provide the results shown in
the next Table:

{\footnotesize\begin{center}
\begin{tabular}{|c|c|} \\[-0.6cm] \hline
$\sigma$ & $\Delta M({\rm TK1})/\hbar m$ \\ \hline $0.5$ & $
-0.962386 $ \\ $0.6$ & $ -0.970537 $ \\ $0.7$ & $ -0.981183 $ \\
$0.8$ & $ -0.994487 $ \\ $0.9$ & $ -1.01053 $ \\ $1.0$ & $ -1.0293
$ \\ \hline
\end{tabular} \hspace{0.1cm}
\begin{tabular}{|c|c|} \\[-0.6cm] \hline
$\sigma$ & $\Delta M({\rm TK1})/\hbar m$ \\ \hline $1.1$ & $
-1.05073 $ \\ $1.2$ & $ -1.07468$ \\ $1.3$ & $ -1.10097 $ \\ $1.4$
& $ -1.12939 $ \\ $1.5$ & $ -1.15971 $ \\ $1.6$ & $ -1.19174 $ \\
\hline
\end{tabular} \hspace{0.1cm}
\begin{tabular}{|c|c|} \\[-0.6cm] \hline
$\sigma$ & $\Delta M({\rm TK1})/\hbar m$ \\ \hline $1.7$ & $
-1.22526 $ \\ $1.8$ & $ -1.2599 $ \\ $1.9$ & $ -1.29571 $ \\ $2.0$
& $ -1.33324 $  \\ $2.1$ & $ -1.37074 $ \\ $2.2$ & $ -1.41007 $ \\
\hline
\end{tabular}
\end{center}}

In Figure 3 the results are plotted as (white) squares and are
compared with the data obtained from the DHN procedure, (black)
dots. We can check the concordance between the two methods; the
relative error for the answer from the asymptotic method is, for
instance, 0.79 \% for $\sigma=1.5$, 0.055 \% for $\sigma=2.0$ and
0.004 \% for $\sigma=2.2$. The greater $\sigma$, the more exact
the response reached from (\ref{eq:as}). On the other hand, for
the range $\sigma \in (0,1)$ the answer is less precise  (for
$\sigma=0.9$ the relative error is 6.0 \%). The reason for this
becomes clear taking into account that the greater the value of
the masses, the smaller the value of
$\frac{\gamma[n-1,v_a^2]}{v_a^{2n-2}}$,  and the more rapidly the
series expansion in formula (\ref{eq:as}) converges.

\begin{figure}[htbp]
\centerline{\epsfig{file=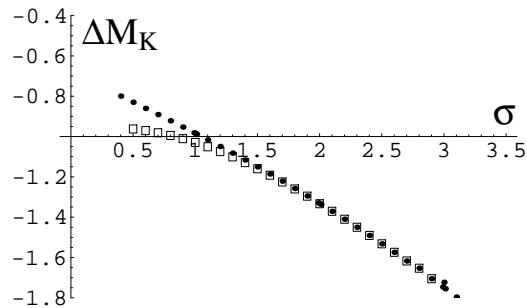,height=4cm}} \caption{\small
{\it One-loop correction to the one-component topological kink
(TK1) mass in units of $\hbar m$. $\bullet$,  DHN formula .
$\Box$, asymptotic series.}}
\end{figure}
It is interesting to note that the answer obtained by means of the
asymptotic series is a continuous function of $\sigma$. This
method, based in the heat kernel expansion and the associated
generalized zeta function, leads directly to the correct result,
without needing to carefully balance the contributions from the
discrete and the continuous parts of the spectrum. The reason is
that the Seeley coefficients are defined in terms of the potential
(and its derivatives), which encodes all the properties of the
spectrum.

\section{Semi-classical masses of kink families}

So far we have only computed the one-loop correction to the
classical mass for the TK1 kink- using either the (exact) DHN
formula (\ref{eq:dhn}) or the (approximate) asymptotic expansion
(\ref{eq:as}). The model, however, exhibits a continuous family of
kinks which have the same classical energy. Our goal in this
Section is to compute the one-loop correction to the kink mass for
each member of the family and to analyze the fate of the kink
degeneracy in the semi-classical level, using formula
(\ref{eq:as}) with the help of some numerical analysis.

\subsection{$\sigma=2$}

In this case, it is possible to find explicit analytical
expressions for the solutions corresponding to the kink orbits
(\ref{eq:tra1}), see \cite{Aai3}:
\begin{equation}
\vec{\phi}_{TK2}[x;a,b]=\frac{(-1)^\alpha }{2} \frac{\sinh 2
(x+a)}{\cosh 2 (x+a)+b}\vec{e}_1+ \frac{(-1)^\beta}{2}
\frac{\sqrt{b^2-1}}{\cosh 2 (x+a)+b}\vec{e}_2 \qquad ,
\label{eq:fam2}
\end{equation}
where $a\in {\Bbb R}$, $b=-\frac{c}{\sqrt{c^2-16}}\in (1,\infty)$,
and $\alpha ,\beta \in {\Bbb Z}/{\Bbb Z}_2$. The family of
Schr\"odinger operators ruling the small fluctuations in the
original fields is in this case
\[
{\cal K}(b)= \left( \begin{array}{cc}
-\frac{d^2}{dx^2}+6\frac{\sinh^2(2x)+b^2-1}{(\cosh(2x)+b)^2}-2 &
12\sqrt{b^2-1}\frac{\sinh(2x)}{(\cosh(2x)+b)^2}
\\[0.2cm]
 12\sqrt{b^2-1}\frac{\sinh(2x)}{(\cosh(2x)+b)^2} &
-\frac{d^2}{dx^2}+6\frac{\sinh^2(2x)+b^2-1}{(\cosh(2x)+b)^2}-2\end{array}
\right) \qquad .
\]
The asymptotic method can be applied by substituting the
corresponding expressions for this case in formula (\ref{eq:as}):
$j=2$, $v_a^2=4$, $a=1,2$, and the potential terms of the above
operator are required in the definitions of the Seeley
coefficients $[a_n]_{aa}({\cal K})$. This calculation has been
carried out for some kinks of the family (\ref{eq:fam2}) with
integration constant $c$ in the range $c\in [-30,c^S)$, see Table
and Figure 4. In sum, the approximate value of the one-loop
quantum mass correction to all these solutions is $-1.33280 \pm
0.00001$ in units of $\hbar m$. Thus, the statement that the
degeneracy of the classical mass to the kink family in the case
$\sigma=2$ is preserved in the quantum framework is highly
accurate.

For this value of $\sigma$ we can easily prove that the statement
is completely exact. A rotation of $45^0$ in ${\Bbb R}^2$,
$\vec{e}_1=\frac{1}{\sqrt{2}}(\vec{\varepsilon}_1+\vec{\varepsilon}_2)$,
$\vec{e}_2=\frac{1}{\sqrt{2}}(\vec{\varepsilon}_1-\vec{\varepsilon}_2)$,
shows that the system is non-coupled. Writing
$\vec{\phi}=\psi_1\vec{\varepsilon}_1+\psi_2\vec{\varepsilon}_2$,
we have that:
\[
T_{\sigma=2}=\frac{1}{2}
\left(\frac{d\psi_1}{dx}\right)^2+\frac{1}{2}\left(\frac{d\psi_2}{dx}\right)^2
\qquad , \qquad
U_{\sigma=2}=4\left(\psi_1^2-\frac{1}{8}\right)+4\left(\psi_2^2-\frac{1}{8}\right)\qquad
.
\]
We can write the degenerate kink family in the alternative form:
\[
\vec{\phi}_{TK2^{*}}[x;a_1,a_2]=\frac{(-1)^\alpha }{2\sqrt{2}}
\tanh(x+a_1)\vec{\varepsilon}_1+ \frac{(-1)^\beta}{2\sqrt{2}}
\tanh(x+a_2)\vec{\varepsilon}_2 \quad ,
\]
in terms of the new parameters $a_1,a_2\in (-\infty,\infty)$. Note
that the TK1 kinks correspond to $a_1=a_2$. In these variables the
Hessian operator is diagonal for any member of the TK2 family:
\[
{\cal K}(a_1,a_2)= \left( \begin{array}{cc}
-\frac{d^2}{dx^2}+4-\frac{6}{\cosh^2(x+a_1)} & 0
\\ 0 &
-\frac{d^2}{dx^2}+4-\frac{6}{\cosh^2(x+a_2)}\end{array}
\right)\qquad .
\]
Therefore, both the spectrum of ${\cal K}(a_1,a_2)$ and the
one-loop correction to the kink masses are independent of
$a_1,a_2\in (-\infty,\infty)$:  $\Delta M(TK2^{*}[a_1,a_2])=\hbar
m(\frac{1}{\sqrt{3}}-\frac{6}{\pi})\approx -1.33281$. The kink
degeneracy is not broken by quantum fluctuations at the one-loop
level; this result would be expected from general arguments
because the family of second-order fluctuation operators above is
isospectral.

\subsection{Numerical approximation}

The problem that we face is the lack of explicit analytical
expressions of the kink solutions for generic values of $\sigma$;
only the kink orbits (\ref{eq:tra1}) are available. However we can
solve the first-order equations (\ref{eq:bps}) by standard
numerical methods. We set the \lq\lq initial" conditions:
\[
\phi_1(0)=0 \qquad , \qquad \frac{\sigma}{2(1-\sigma)}\phi_2^2(0)
-\frac{c}{2\sigma}|\phi_2(0)|^{\frac{2}{\sigma}}=\frac{1}{4}
\qquad .
\]
The rationale behind this choice is as follows: (1) for any kink
solution, $\phi_1(x)$ has always a zero. Translational invariance
allows us to set the zero at $x=0$. (2) To ensure that we will
find a numerical kink solution, we fix $\phi_2(0)$ in such a way
that (\ref{eq:tra1}) is satisfied for a given value of $\sigma$
and arbitrary choices of $c$.

In this way the numerical method provides us with a succession of
points of the kink solution generated by a interpolation
polynomial. The kink polynomial is then used to calculate the
$[a_n]_{aa}({\cal K})$ coefficients of formula (\ref{eq:as})
approximately and hence the one-loop correction to the kink mass.
A subtle point needs to be clarified: the coefficients are defined
in terms of the field solutions and their derivatives. Whereas it
is all right to describe the kink solutions approximately by
interpolation polynomials, taking \lq\lq derivatives" of such
discrete configurations of points would induce important errors.
Fortunately, we can use the first-order equations (\ref{eq:bps})
to express all the field derivatives as polynomials in the fields.

This procedure was carried out for values of $\sigma$ in the
$(1.3,3.3)$ range with $\Delta \sigma=0.1$ finding in all cases a
similar pattern in the behaviour of the $\Delta M(K(c))$. In the
Figure 4 we show the results obtained using this method for
several values of $c$ in the range $c\in (-30,c^S)$ in the cases
$\sigma=1.5$ and $\sigma=2.5$. A first observation is the perfect
agreement reached between the numerical approximation and the
exact result of the $\sigma=2$ case. Next, we notice that the
greater the value of $\sigma$, the better the convergence of the
asymptotic expansion to the exact value for kink solutions with
$c$ sufficiently distant from $c^S$. We check that the kink
degeneracy is always maintained in a wide range of $c$, but
starts to fail when $c$ approaches $c^S$.

{\footnotesize
\begin{center}
{\begin{tabular}{cc} $\sigma=1.5$ & \\ \hline $c$ & $\Delta M$ \\
\hline $-30$ & $-1.16009$ \\ $-27.5$ & $-1.16017$ \\ $-25$ &
$-1.16128$ \\ $-22.5$ & $-1.16042$ \\ $-20$ & $-1.16061$ \\
$-17.5$ & $-1.16088$ \\ $-15$ & $-1.16128$ \\ $-12.5$ & $-1.16193$
\\ $-10$ & $-1.16313$ \\ $-7.5$ & $-1.16597$ \\ $-5$  & $-1.18205$
\\ $-4.6801886$ & $-1.24345$ \\ $-4.68018860186678332$ &
$-1.25103$ \\ \hline
 & \\
 & \\
\end{tabular}} \hspace{0.2cm}
{\begin{tabular}{cc} $\sigma=2.0$ & \\ \hline $c$ & $\Delta M$ \\
\hline $-30$ & $-1.33281$ \\ $-27.5$ & $-1.33281$ \\ $-25$ &
$-1.33281$ \\ $-22.5$ & $-1.33281$ \\ $-20$ & $-1.33281$ \\
$-17.5$ & $-1.33281$ \\ $-15$ & $-1.33281$ \\ $-12.5$ & $-1.33281$
\\ $-10$ & $-1.33281$ \\ $-7.5$ & $-1.33281$ \\ $-5$  & $-1.33280$
\\ $-4.001$  & $-1.33280$ \\ $-4.00001$  & $-1.33280$ \\ \hline &
\\ & \\
\end{tabular}}
\hspace{0.2cm} {\begin{tabular}{cc} $\sigma=2.5$ & \\ \hline
 $c$ & $\Delta M$ \\ \hline
$-30$ & $-1.52784$ \\ $-27.5$ & $-1.52782$ \\ $-25$ & $-1.52780$
\\ $-22.5$ & $-1.52778$ \\ $-20$ & $-1.52774$ \\ $-17.5$ &
$-1.52769$
\\ $-15$ & $-1.52760$ \\ $-12.5$ & $-1.52744$ \\ $-10$ &
$-1.52711$ \\ $-7.5$ & $-1.52626$ \\ $-5$  & $-1.52285$ \\ $-4$  &
$-1.52168$ \\ $-3.97$ & $-1.52915$ \\ $-3.96594571$ & $-1.55402$
\\ $-3.96594570565808127$ & $-1.56127$ \\ \hline
\end{tabular}}
\end{center}}

\noindent\begin{figure}[htbp] \centerline{
\epsfig{file=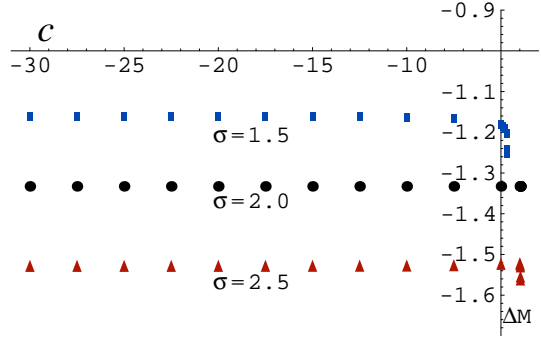,height=4.5cm}} \caption{\small \it The
One-loop Quantum Mass Correction in the cases $\sigma=1.5$,
$\sigma=2.0$ and $\sigma=2.5$. }
\end{figure}

Thus, the previous Table and Figure suggest the survival of kink
degeneracy at the one-loop level up to values of $c$ close to
$c^S$. A more precise idea about what is going on in the vicinity
of $c^S$ is given by the following observations:

1) We first notice that the $c=c^S$ kink orbits belong to the
${\cal C}_{(a,b)}^{\pm\mp} (a\neq b$) topological sectors for any
value of $\sigma$. Thus, this kind of kink is called a \lq\lq
link" because the associated orbits link different points in the
vacuum moduli space. The trajectories of these link kinks enclose
the orbits of the kink family studied previously in Section \S 4.
For $\sigma=2$, these TK2L topological kinks of link type are:
\[
\vec{\phi}_{\rm TK2L}(x)= \frac{(-1)^\alpha}{4}(1-(-1)^\gamma
\tanh(x+a))\vec{e}_1+ \frac{(-1)^\beta}{4} (1+(-1)^\gamma\tanh
(x+a))\vec{e}_2
\]
with $a\in {\Bbb R}$, $\alpha, \beta, \gamma \in {\Bbb Z}/{\Bbb
Z}_2$. The link topological kinks for $\sigma=\frac{1}{2}$ are
\[
\vec{\phi}_{\rm TK2L}(x)= \frac{(-1)^\alpha}{4}\left(1-(-1)^\beta
\tanh \frac{x+a}{2}\right)\vec{e}_1+ \frac{(-1)^\gamma}{\sqrt{2}}
\sqrt{1 +(-1)^\beta \tanh \frac{x+a}{2}}\vec{e}_2
\]
with $\alpha, \beta, \gamma \in {\Bbb Z}$. There are no analytical
expressions available for other values of $\sigma$.

2) Second, we realize that because the existence of these
enveloping kinks the energy density of any kink obtained for $c$
near $c^S$ is formed by two lumps. We explain the general
situation in the $\sigma=\frac{1}{2}$ case. Even though the
asymptotic method works poorly for this value of $\sigma$ (with
errors higher than a 18\%), the analytical information allows a
full description of the situation. The TK2 kink family is given
by:
\[
\vec{\phi}_{TK2}[x;b]=\frac{(-1)^\alpha }{2} \frac{\sinh
(x+a)}{\cosh (x+a)+b} \vec{e}_1 +(-1)^\beta \sqrt{\frac{b}{\cosh
(x+a)+b}}\vec{e}_2
\]
where $a\in {\Bbb R}$, $b=\frac{1}{\sqrt{1-4c}}\in (0,\infty)$,
and , $\alpha ,\beta \in {\Bbb Z}/{\Bbb Z}_2$.

In Reference \cite{Aai1}, we showed that the splitting into two
lumps starts at $c=0$ or $b=1$.  In the previous subsection we
have observed through numerical computations for any value of
$\sigma\neq 2$ that the departure from the TK1 quantum correction
also starts at the value of $c$ where the splitting begins. Use of
the $c$ parameter is necessary to implement the numerical method,
but unsuitable for discussing this phenomenon. Note that $c\in
(0,c^s)$ is tantamount to $b\in (1,+\infty )$. $b$ is a measure of
the distance between the two lumps, although a highly non-linear
one, see \cite{Aai1}. Thus, the breaking of the degeneracy is
noticeable when the two lumps are further apart. The induction of
this repulsive force by quantum effects can be better understood
by looking at the family of the TK2 Hessian operators:
\[
{\cal K}(b)= \left( \begin{array}{cc}
-\frac{d^2}{dx^2}+\frac{3b}{\cosh x+b}+\frac{6\sinh^2x}{(\cosh
x+b)^2}-2 & \frac{6\sqrt{b}\sinh x}{(\cosh x+b)^{\frac{3}{2}}}
\\\frac{ 6\sqrt{b}\sinh x}{(\cosh x+b)^{\frac{3}{2}}} &
-\frac{d^2}{dx^2}+\frac{3}{2}\frac{b}{\cosh x+b}+\frac{3}{4}
\frac{\sinh^2x}{(\cosh x+b)^2}-\frac{1}{2}\end{array}
\right)\qquad .
\]
In the Figure 5 we show a plot of the diagonal components of the
potential for several values of $c$.

\begin{figure}[htbp] \centerline{
\epsfig{file=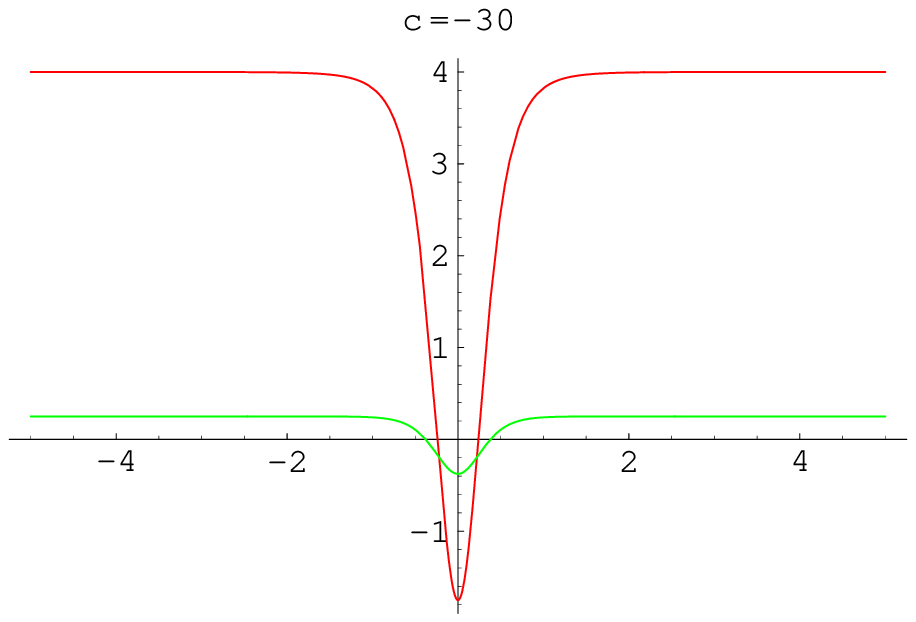,height=3.5cm}\quad
\epsfig{file=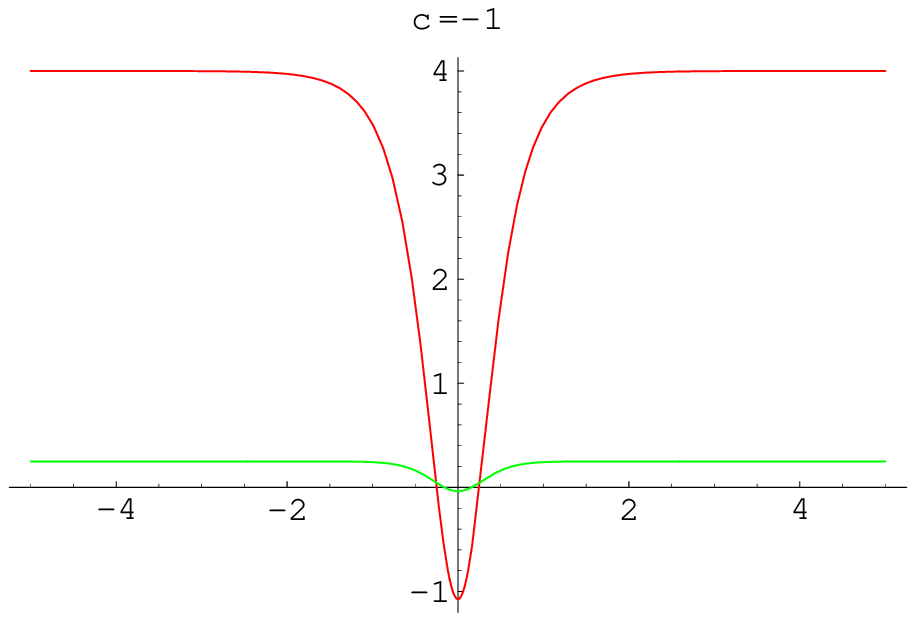,height=3.5cm}}
\centerline{\epsfig{file=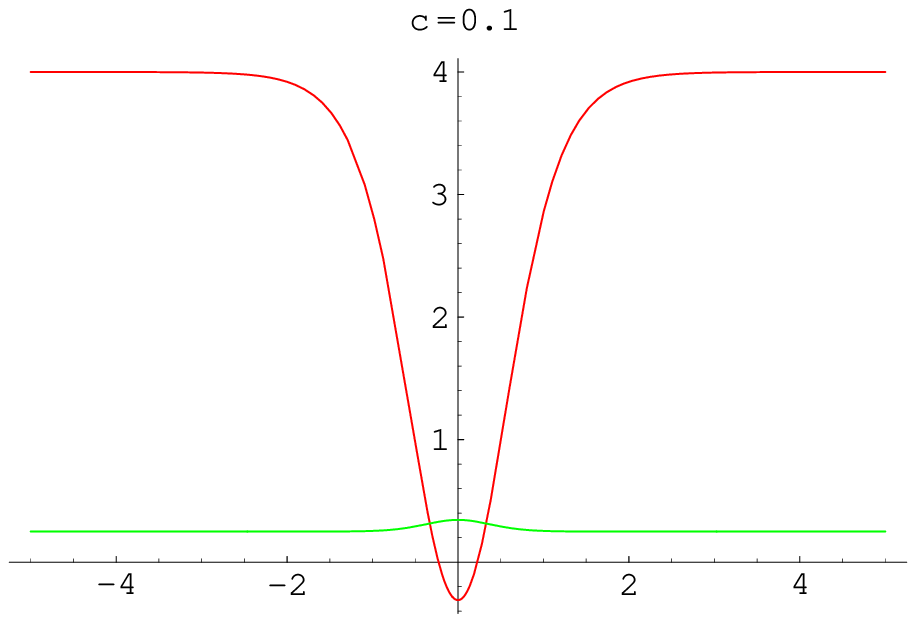,height=3.5cm}\quad
\epsfig{file=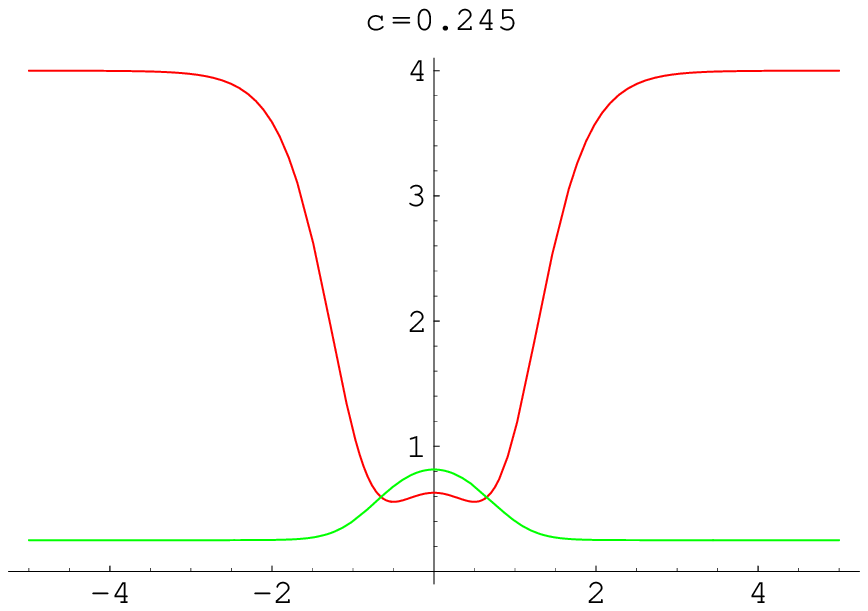,height=3.5cm}\quad
\epsfig{file=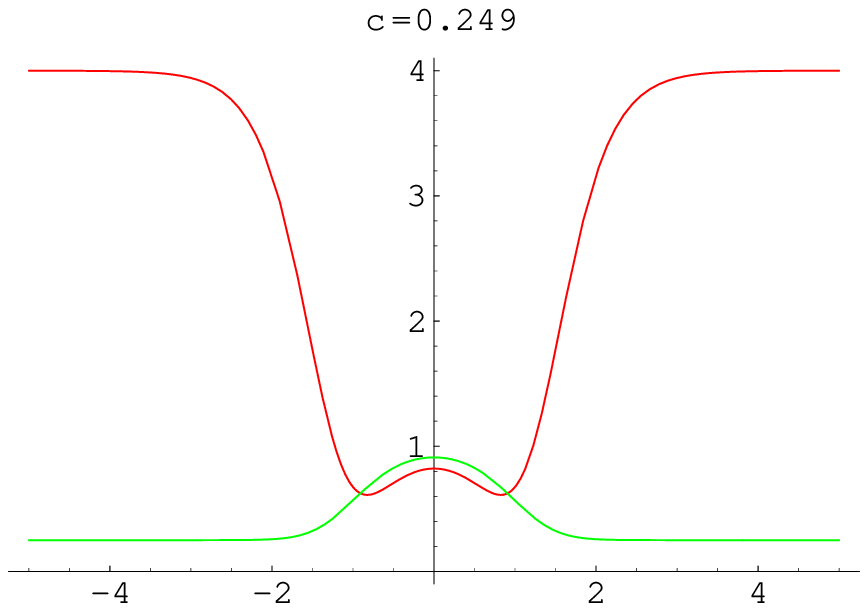,height=3.5cm}} \caption{\small \it
Diagonal components of the potential for c=-30, c=-1, c=0.1,
c=0.245 and c=0.249.}
\end{figure}

There is no modification of the potential between $c=-30$ and
$c=-1$ (the off-diagonal components are odd functions and the area
enclosed by them is zero); there must be kink degeneracy in this
range of $c$. Starting at $c=0$, the second component turns from
potential well to barrier, inducing a (still) very weak repulsion.
The closer the value of $c$ to $c^S=\frac{1}{4}$ , the stronger
the repulsion coming from the second component, whereas the first
component develops a double well, clearly arising from the meiosis
into two lumps. Therefore, the spectrum of ${\cal K}(b)$ is
completely different in the regimes $b\in (0,1)$ and $b\in
(10^5,\infty)$, with a transition region in between. We stress
again that the picture is similar to this for any $\sigma \neq 2$:
the classical degeneracy starts to fail for the value of $c$ where
two peaks in the energy density appear: quantum fluctuations
induce repulsive forces between the two basic lumps, which are
constituents of a TK2 kink. The data in the Table also show that
the rate of change of the quantum correction diminishes when the
two lumps are extremely far apart, in a range of $c$ where there
are no significant changes in the structure of ${\cal K}(c)$.

We stress that the breaking of the kink degeneracy due to quantum
effects suggests the possibility of a similar phenomenon between
higher dimensional topological defects as vortices and monopoles;
it is tantalizing to think that the method here developed could
also be applied to study the one-loop effects on the moduli spaces
of vortices and monopoles. Nevertheless, one can argue the
survival of kink degeneracy to quantum fluctuations in the fully
supersymmetric system. The reason is the saturation of the BPS
bound at the quantum level in ${\cal N}=1$ Wess-Zumino $d=2$
supersymmetric models due to equal anomalies in the energy and the
central charge; see Reference \cite{Fuj} for a recent and elegant
proof in systems with only one chiral superfield. In our case, two
chiral superfields, the anomaly is given - up to some constant- in
terms of the Laplacian of the superpotential:
\[
|\Delta W(\vec{\phi}(\infty))-\Delta
W(\vec{\phi}(-\infty))|=4+2\sigma \qquad ,
\]
a quantity independent of c. Therefore, the c-dependence in the
energy coming from bosonic fluctuations must be exactly cancelled
by taking into account the fermionic fluctuations with SUSY
preserving boundary conditions.

We end this Section by exploring some points about the one-loop
correction to the TK2L link kinks. If $\sigma=2$, the TK2L
Hessian/Schrodinger operator for $a=0$ is
\[
{\cal K}^L(\sigma=2)= \left( \begin{array}{cc}
-\frac{d^2}{dx^2}+4-\frac{3}{\cosh^2x} & \frac{3}{\cosh^2x}
\\\frac{3}{\cosh^2x} &
-\frac{d^2}{dx^2}+4-\frac{3}{\cosh^2x}\end{array} \right)\qquad ,
\]
whereas if $\sigma=\frac{1}{2}$, also for $a=0$, we have:
\[
{\cal K}^L(\sigma=\frac{1}{2})= \left( \begin{array}{cc}
-\frac{d^2}{dx^2}+\frac{5}{2}-\frac{3}{2}\tanh(\frac{x}{2})-
\frac{3}{2}\frac{1}{\cosh^2(\frac{x}{2})} &
\frac{3}{2\sqrt{2}}(1-\tanh(\frac{x}{2}))\sqrt{1+\tanh(\frac{x}{2})}
\\\frac{3}{2\sqrt{2}}(1-\tanh(\frac{x}{2}))\sqrt{1+\tanh(\frac{x}{2})}
 &
-\frac{d^2}{dx^2}+\frac{5}{8}+\frac{3}{8}\tanh(\frac{x}{2})-
\frac{3}{16}\frac{1}{\cosh^2(\frac{x}{2})} \end{array}
\right)\qquad .
\]
In this last case, we would need to use the background
renormalization method as developed in \cite{Aai0} to compute the
one-loop correction, but the task is so involved that we only
discuss the $\sigma=2$ case.

It is clear that ${\rm Spec}\,{\cal K}^L(\sigma=2)={\rm
Spec}\,{\cal D}$, where:
\[
{\cal D}= \left( \begin{array}{cc} -\frac{d^2}{dx^2}+4 &  0
\\ 0 &
-\frac{d^2}{dx^2}+4-\frac{6}{\cosh^2x}\end{array} \right)\qquad .
\]
One immediately concludes that
\[
\Delta M(TK2L)=\frac{1}{2}\Delta M(TK1) \quad ,
\]
i.e. the famous kink sum rule also holds at the one-loop level.
The results obtained via the asymptotic expansion agree with this
observation. The very accurately calculated degeneracy in the case
$\sigma=2$, obtained from numerical plus asymptotic computations
even in the vicinity of the limiting value $c^S=-4$ (see Figure
4), implies that the combination of two identical link kinks
(reached when the parameter $c$ goes to $c^S$) has the same mass
quantum correction as the rest of the kinks.

As we have seen the situation is different (and to some extent
surprising) for other values of $\sigma$. The singular behaviour
of the system for $\sigma=2$ is related to the fact that the
metric ruling the kink adiabatic motion is Euclidean if
$\sigma=2$, see \cite{Aai1}; consequently, the two lumps move
freely with respect to each other.

\section{Classical versus semi-classical kink degeneracy}

In this Section we shall address the subtle question of what
happens to the kink degeneracy when quantum effects are taken into
account from the analytical - more than the computational- side.
Let us denote by
\[
\vec{\phi}^{K(c)}(x;c)={\bar\phi}_1(x;c)\vec{e}_1+{\bar\phi}_2(x;c)\vec{e}_2
\]
the solution of the first-order equations (\ref{eq:bps}) given by
the orbit (\ref{eq:tra1}). Let us define ${\cal K}(c)={\cal
V}_{1}-V(c)$ as:
\[
{\cal K}
(c)=\left(\begin{array}{cc}-\frac{d^2}{dx^2}+24{\bar\phi}_1^2[x;c]+4\sigma(\sigma+1)
{\bar\phi}_2^2[x;c]-2 &
8\sigma(\sigma+1){\bar\phi}_1{\bar\phi}_2[x;c]\\
8\sigma(\sigma+1){\bar\phi}_1{\bar\phi}_2[x;c] &
-\frac{d^2}{dx^2}+4\sigma(\sigma+1){\bar\phi}_1^2[x;c]+6\sigma^2
{\bar\phi}_2^2[x;c]-\sigma\end{array}\right)\qquad .
\]
We have thus a family of Schrodinger operators governing the small
fluctuations around any kink in the degenerate family.

The one-loop correction to the mass of the $K(c)$ kink is given by
the formula (5) of Reference \cite{Aai2}:
\begin{eqnarray}
\Delta M(K(c))&=&\lim_{s\rightarrow
-\frac{1}{2}}[\Delta_1\varepsilon^{K(c)}(s)+\Delta_2\varepsilon^{K(c)}(s)]\nonumber
\\
\Delta_1\varepsilon^{K(c)}(s)&=&\frac{\hbar}{2}\mu^{2s+1}\left[\zeta_{P{\cal
K}(c)}(s)-\zeta_{{\cal V}_{1}}(s)\right]\nonumber \\
\Delta_2\varepsilon^{K(c)}(s)&=&\lim_{L\rightarrow\infty}\frac{\hbar}{2L}\mu^{2s+1}\frac{\Gamma(s+1)}{\Gamma(s)}
\sum_{a=1}^2\zeta_{({\cal
V}_1)_{aa}}(s+1)\int_{-\frac{mL}{2}}^{\frac{mL}{2}} dx\,
V_{aa}(x;c) \qquad .
\end{eqnarray}
We recall that $\Delta_1\varepsilon^{K(c)}(-{1\over 2})$ is the
kink Casimir energy measured with respect to the vacuum Casimir
energy. $\Delta_2\varepsilon^{K(c)}(-{1\over 2})$, however,
accounts for the contribution to the kink energy of the  mass
renormalization counterterm.

The generalized zeta functions of ${\cal K}(c)$ (with the zero
eigenvalues excluded) and ${\cal V}_1$ can be written in terms of
the heat functions via Mellin transforms:
\[
\zeta_{P{\cal K}(c)}(s)=\frac{1}{\Gamma (s)}\int_0^\infty\ d\beta
\beta^{s-1}{\rm Tr}e^{-\beta P{\cal K}(c)}\qquad , \qquad
\zeta_{{\cal V}_1}(s)=\frac{1}{\Gamma (s)}\int_0^\infty\ d\beta
\beta^{s-1}{\rm Tr}e^{-\beta {\cal V}_1}\quad .
\]
The derivative of $\Delta_1\varepsilon^{K(c)}(s)$ with respect to
$c$ is obtained immediately,
\[
\frac{\partial}{\partial
c}\Delta_1\varepsilon^{K(c)}(s)=\frac{\hbar}{2\Gamma(s)}\mu^{2s+1}\int_0^\infty\
d\beta \beta^{s-1}{\rm Tr}\left[\frac{\partial V}{\partial
c}e^{-\beta P{\cal K}(c)}\right]\quad ,
\]
whereas we also have:
\[
\frac{\partial}{\partial c}
\Delta_2\varepsilon^{K(c)}(s)=\lim_{L\rightarrow\infty}\frac{\hbar}{2L}\mu^{2s+1}
\frac{\Gamma(s+1)}{\Gamma(s)} \sum_{a=1}^2\zeta_{ ({\cal
V}_1)_{aa}}(s+1)\int_{-\frac{mL}{2}}^{\frac{mL}{2}} dx\,
\frac{\partial V_{aa}}{\partial c} \quad .
\]

If the sum of these two quantities adds to zero at the
$s\rightarrow -\frac{1}{2}$ limit we can be sure that the kink
degeneracy also holds up to this order in the $\hbar$-expansion.
Unfortunately, an exact computation is beyond our analytical
capacities but we can work an asymptotic expansion of this formula
along similar lines to those used in Reference \cite{Aai2} and
\cite{Avra}. From the high-temperature expansion of the heat
kernel,
\begin{displaymath}
K_{{\cal K}(c){ab}}(x,x;\beta)=\sum_{d=1}^2
A_{ad}(x,x;\beta)\frac{e^{-\beta
v_d^2}}{\sqrt{4\pi\beta}}\delta_{db}=\frac{e^{-\beta
v_b^2}}{\sqrt{4\pi\beta}}\sum_{n=0}^\infty[a_n]_{ab}(x,x)\beta^n
\end{displaymath}
and bearing in mind that
\begin{eqnarray*}
{\rm Tr}\left[\frac{\partial V}{\partial c}e^{-\beta P{\cal
K}(c)}\right]&=&\int_{-\infty}^\infty dx\, {\rm
tr}\left<x\right|\frac{\partial V}{\partial c}e^{-\beta P{\cal
K}(c)}\left|x\right>\\&=& \int_{-\infty}^\infty dx
\int_{-\infty}^\infty dx^\prime\, {\rm
tr}\left<x\right|\frac{\partial V}{\partial
c}\left|x^\prime\right>\left<x^\prime\right|e^{-\beta P{\cal
K}(c)}\left|x\right>\\&=&\sum_{a,b=1}^2\int_{-\infty}^\infty dx
\frac{\partial V_{ab}(x;c)}{\partial c}K_{{\cal
K}(c){ba}}(x,x;\beta)
\end{eqnarray*}
we find
\begin{eqnarray*}
{\rm Tr}[\frac{\partial V}{\partial c}e^{-\beta {\cal
K}(c)}]&=&\frac{1}{\sqrt{4\pi\beta}}\sum_{n=0}^\infty\sum_{a,b=1}^2\int_{-\infty}^\infty
dx\, \left( \frac{\partial V_{ab}(x)}{\partial
c}[a_n]_{ba}(x,x)e^{-\beta
v_a^2}\right)\beta^n\\&=&\frac{1}{\sqrt{4\pi}}\sum_{n=0}^\infty\sum_{a=1}^2
[c_n]_a e^{-\beta v_a^2} \beta^{n-\frac{1}{2}}\qquad ,
\end{eqnarray*}
where
\begin{displaymath}
[c_n]_a=\sum_{b=1}^2\int_{-\infty}^\infty dx\, \frac{\partial
V_{ab}(x;c)}{\partial c}[a_n]_{ab}(x,x)\qquad .
\end{displaymath}
Therefore,
\begin{displaymath}
\frac{\partial}{\partial c}\zeta_{{\cal
K}(c)}(s)=\frac{1}{\sqrt{4\pi}\Gamma[s]}\sum_{n=0}^\infty\sum_{a=1}^2[c_n]_a\int_0^1
d \beta e^{-\beta v_a^2}\beta^{s+n-\frac{1}{2}}+\frac{\partial
B_{{\cal K}(c)}(s)}{\partial c}\qquad ,
\end{displaymath}
where $B_{{\cal K}(c)}(s)$ is a negligible error arising from the
truncation of the $\beta$ integration in $\beta=1$. This gives:
\begin{displaymath}
\frac{\partial}{\partial c}\Delta_1
\varepsilon^{K(c)}(s)=\frac{\hbar m}{2}\mu^{2s+1}\sum_{n=0}^\infty
\sum_{a=1}^2\frac{[c_n]_a\gamma[s+n+\frac{1}{2},v_a^2]}{\sqrt{4\pi}\,v_a^{2(s+n+\frac{1}{2})}}
+{\rm error}\qquad .
\end{displaymath}
Also, we have that
\begin{displaymath}
\zeta_{({\cal V}_1)_{aa}}(s+1)=\frac{mL}{\Gamma[s+1]}
\frac{\gamma[s+\frac{1}{2},v_a^2]}{\sqrt{4\pi}\,v_a^2(s+\frac{1}{2})}
\end{displaymath}
and, because $[a_0]_{ab}(y,y)=\delta_{ab}$,
\[
[c_0]_a=\sum_{b=1}^2\int_{-\infty}^\infty dy\, \frac{\partial
V_{ab}}{\partial c}\qquad .
\]
Thus,
\begin{displaymath}
\frac{\partial}{\partial c
}\Delta_2\varepsilon^{K(c)}(s)=-\frac{\hbar m
}{2}\mu^{2s+1}\sum_{a=1}^2
\frac{\gamma[s+\frac{1}{2},v_a^2]}{\sqrt{4\pi}v_a^2(s+\frac{1}{2})}[c_0]_a
\end{displaymath}
and finally we obtain:
\begin{equation}
\frac{\partial}{\partial c}\left[\Delta_1 \varepsilon^{K(c)}(s)+
\Delta_2 \varepsilon^{K(c)}(s)\right]=\frac{\hbar m}{2}\mu^{2s+1}
\sum_{n=1}^\infty\sum_{a=1}^2\frac{[c_n]_a\gamma[s+n+\frac{1}{2},v_a^2]}{\sqrt{4\pi}\,
v_a^{2(s+n+\frac{1}{2})}}+{\rm error}  \quad . \label{eq:derq}
\end{equation}
It is a remarkable fact that the pole arising in
$\lim_{s\rightarrow -\frac{1}{2}}\Delta_1 \varepsilon^{K(c)}(s)$
appears in the $n=0$ term of the asymptotic expansion. The residue
takes a value such that this divergence is exactly cancelled by
$\lim_{s\rightarrow -\frac{1}{2}}\Delta_2 \varepsilon^{K(c)}(s)$.
Explicit computation of (\ref{eq:derq}) has been carried out for
$\sigma=2$ and we have perfect concordance with the data depicted
in the Figure 5. In this case, the coefficients $c_n$ vanish,
such that $\frac{\partial}{\partial c} \Delta
\varepsilon^{K(c)}(s)=0$.

\section{Jacobi fields, kink instability and resonances}

In the ${\cal C}_{(2,2)}^{\pm\mp}$ topological sectors, things are
more difficult. As critical points of the energy functional, the
topological defects must satisfy the Euler-Lagrange equations
\begin{equation}
\frac{d^2 \phi_1}{d x^2} = 2 \phi_1 \left(4 \phi_1^2 + 2 \sigma
(1+\sigma) \phi_2^2-1\right) \hspace{1.2cm} \frac{d^2 \phi_2}{d
x^2} = \sigma \phi_2 \left( 4(\sigma+1) \phi_1^2 + 2 \sigma
\phi_2^2-1\right)\qquad .  \label{eq:ecuseg}
\end{equation}
If $\sigma=2$ and $\sigma=\frac{1}{2}$, we are able to obtain all
the kinks in these sectors of the model because the associated
dynamical problem is integrable.
\begin{itemize}
\item For $\sigma=2$ in the ${\cal C}_{(2,2)}^{\pm\mp}$ topological sectors, we find
\[
\vec{\phi}_{TK2'}[x;b]= \frac{(-1)^\alpha}{2}
\frac{\sqrt{b^2-1}}{\cosh 2 (x+a)+b} \vec{e}_1+
\frac{(-1)^\beta}{2} \frac{\sinh 2 (x+a)}{\cosh 2
(x+a)+b}\vec{e}_2 \qquad .
\]
\item For $\sigma=\frac{1}{2}$, in the ${\cal C}_{(2,2)}^{\pm\mp}$ topological sectors
we have the family of solutions:
\begin{eqnarray*}
\vec{\phi}_{TK2'}[x;b]&=&(-1)^\alpha\frac{\sinh  b \sinh
(x+a)}{\cosh^2 (x+a)+2 \cosh b\cosh (x+a)+1}\vec{e}_1 +\\
&+&(-1)^\beta\frac{-\sinh (x+a)}{\sqrt{\cosh^2 (x+a)+2 \cosh b
\cosh (x+a)+1}}\vec{e}_2 \qquad .
\end{eqnarray*}
\end{itemize}

For generic $\sigma$, only the TK1' kinks are known explicitly in
${\cal C}_{(2,2)}^{\pm\mp}$:
\[
\vec{\phi}_{{\rm TK1'}}(x)=(-1)^\alpha\frac{1}{\sqrt{2\sigma}}{\rm
tanh}\sqrt{\frac{\sigma}{2}}(x+a)\qquad .
\]
Defining $z=\sqrt{\frac{\sigma}{2}}x$, we write the Hessian for
the TK1' kinks as
\begin{equation} {\cal J}=\frac{\sigma}{2}\left( \begin{array}{cc}
-\frac{d^2}{dz^2}+4 -\frac{4(1+\sigma)}{\sigma} \sech^2 z & 0 \\ 0
& -\frac{d^2}{dz^2}+4 -6\sech^2 z
\end{array} \right) \quad . \label{eq:hesstk11}
\end{equation}
${\cal J}$ is a diagonal matrix differential operator that
includes as the ${\cal J}_{22}$ matrix element the very well known
Posch-Teller Schr\"odinger operator, with $c_0^2=4$ and $u_0=6$.
${\cal J}_{11}$ is also of Posch-Teller type - $c_0^2=4$,
$u_0=\frac{4(1+\sigma)}{\sigma}$-, and deserves a closer analysis.
The eigenvalues $\omega_n=4-[\kappa-(n+\frac{1}{2})]^2$,
$\kappa=\sqrt{\frac{4(1+\sigma)}{\sigma}+\frac{1}{4}}$,
$n=0,1,2,\cdots , I[\kappa -\frac{1}{2}]-1$ belong to the discrete
spectrum of ${\cal J}_{11}$. Thus,
\[
{\rm Spec}{\cal J}= \cup_{n=0}^{I[\kappa-\frac{1}{2}]-1}
\left\{4-[\kappa-(n+\textstyle\frac{1}{2})]^2\right\}\cup
\{k_1^2+4\} \cup \{0\}\cup \{3\}\cup \{k_2^2+4\}_{k_2\in {\Bbb
R}}\qquad .
\]
The phase shifts for the continuous spectrum are
\[
\delta(k_1)=\delta_{+}(k_1)+\delta_{-}
(k_1)\hspace{0.1cm};\hspace{1cm}\delta (k_2)=-2{\rm
arctan}\frac{3k_2}{2-k_2^2} \qquad .
\]
The reflection coefficient associated to the Schr\"odinger
operator ${\cal J}_{11}$ is not zero if
$\sigma\neq\frac{4}{j(j+1)-4}$ with $j$ a positive integer number
greater than or equal to 2. Therefore, there are also in general
even and odd phase shifts,
\[
\delta_{\pm} (k_1)=\frac{1}{4}{\rm arctan}\left(\frac{{\rm
Im}(T(k_1)\pm R(k_1))}{{\rm Re}(T(k_1)\pm R(k_1))}\right) ,
\]
to be read from the transmission and reflection
\[
T(k_1)=\frac{\Gamma(\frac{1}{2}+\kappa-ik_1)\Gamma(\frac{1}{2}-\kappa-ik_1)}{\Gamma(1-ik_1)
\Gamma(-ik_1)}\qquad , \qquad
R(k_1)=\frac{\Gamma(\frac{1}{2}\kappa-ik_1)\Gamma(\frac{1}{2}-\kappa^2-ik_1)\Gamma(ik_1)}
{\Gamma(1+\kappa)\Gamma(1-\kappa^2)\Gamma(-ik_1)}
\]
coefficients.

The physical and geometrical information encoded in the spectrum
of ${\cal J}$ is as follows:
\begin{itemize}
\item There are two bound states in the tangent direction to the
${\rm TK}1'$ kink in ${\cal C}_{(2;2)}^{\pm\mp}$. The first of
them is the translational zero mode and the second one is
interpreted as usual: a scalar boson \lq\lq polarized" in internal
space in the $\vec{e}_2$ direction is captured by the kink. The
scattering eigenstates correspond to $\vec{e}_1$-polarized scalar
bosons moving in the kink background.

\item There are $I[\kappa-\frac{1}{2}]-1$ bound states in the orthogonal
direction to the ${\rm TK}1'$ kink in ${\cal C}_{(2;2)}^{\pm\mp}$.
The associated eigenvalues are positive if $\sigma>2$ and thus the
${\rm TK}1'$ kink is stable in this regime.

\item When $\sigma<2$, there are some negative eigenvalues in ${\rm
Spec}({\cal J})$ and the ${\rm TK}1'$ kink is unstable against the
associated fluctuations. The number of negative eigenvalues and
the degree of instability increase with decreasing $\sigma$: if
$\sigma\in (\frac{4}{j(j-1)-4}, \frac{4}{j(j+1)-4}]$ the spectrum
of ${\cal J}_{11}$ contains $j-2$ negative eigenvalues and there
exist $j-2$ instability directions around the ${\rm TK1'}$ kink in
the ${\cal C}_{( 2;2)}^{\pm\mp}$ sectors. For an analysis of kink
stability in multi-component scalar field theory, see \cite{Aai4}.
\item
At the critical values $\sigma=\frac{4}{j(j+1)-4}$ there is a jump
in the number of negative eigenvalues and the Jacobi field
\[
\psi_J(q)\cong
(1-q^2){}_2F_1[{\textstyle\frac{5}{2}+\sqrt{j(j+1)+\frac{1}{4}}},
{\textstyle\frac{5}{2}-\sqrt{j(j+1)+\frac{1}{4}}},
3,{\textstyle\frac{1}{2}(1+q)}]\hspace{0,3cm};\hspace{0.3cm}q={\rm
tanh }z
\]
becomes a bound state belonging to the discrete spectrum of ${\cal
J}_{11}$. The Morse index theorem, \cite{Aai4}, tells us that
there is a continuous family of TK2' kinks in the ${\cal C}_{(
2;2)}^{\pm\mp}$ sectors if $\sigma=\frac{4}{j(j+1)-4}$. The
members of this family are the flow lines of the gradient of a
second superpotential ${\tilde W}$ that has two critical points at
$\vec{\phi}_{V_2^\pm}=\pm\frac{\sqrt{j(j+1)-4}}{2\sqrt{2}}\vec{e}_2$
and $j-2$ branching points: the zeroes of $\psi_J(x)\vec{e}_2$,
see \cite{Aai4} .

\end{itemize}

\noindent\begin{figure}[htbp] \centerline{\epsfig{file=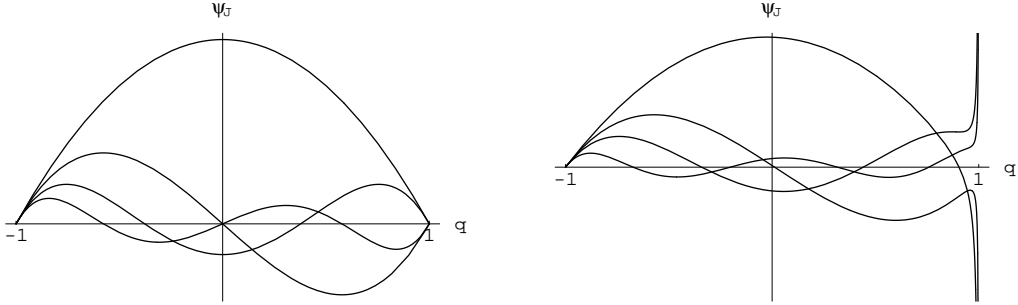,
height=4cm}} \caption{\small \it Graphics of $\psi_J(q)$ for
$j=2,j=3,j=4,j=5$ (left), for non-critical values of $\sigma$
close to the critical ones (right). }
\end{figure}

Therefore, the ${\rm TK}1'$ kink gives rise to a bona fide
eigenstate of the Hamiltonian only if $\sigma>2$. Here we shall
not give $\Delta M({\rm TK}1')$, because the computation is
identical to that developed  for $\Delta M({\rm TK}1)$ if
$\sigma\geq 2$ but gives rise to a complex mass if $\sigma^2<2$;
the TK1' kink state should be interpreted as a resonance in this
regime.

\section*{Appendix: mode number cutoff regularization method}

In this Appendix we present a derivation of the basic equation
(\ref{eq:dhn}) in Section \S 3. Given the ${\cal K}$ and ${\cal
V}$ Schrodinger operators
\[
{\cal V}=-\frac{d^2}{dx^2}+v^2\hspace{0.5cm},\hspace{0.5cm}{\cal
K}=-\frac{d^2}{dx^2}+v^2-V(x)  ,
\]
respectively ruling the small fluctuations around the vacuum and
the kink, the renormalized zero point energy is:
\begin{equation}
\Delta_1\varepsilon^{\rm K}=\frac{\hbar m}{2}\left(\sum_{{\rm
Spec}{\cal K}}\omega_n-\sum_{{\rm Spec}{\cal
V}}{\bar\omega}_n\right) \label{eq:cas} \qquad .
\end{equation}
Here, $v^2=\frac{d^2U}{d\psi^2}|_{\psi_{\rm V}}$ is a constant,
and $v^2-V(x)=\frac{d^2U}{d\psi^2}|_{\psi_{\rm K}}$ gives rise to
a potential well $V(x)$, which rapidly decreases to its asymptotic
value: $\displaystyle\lim_{x\rightarrow\pm\infty}V(x)=0$; note
that, as in the main text, we use dimensionless eigenvalues and
recover the dimension in the final response through the overall
constant of (\ref{eq:cas}). The spectrum of both ${\cal K}$ and
${\cal V}$ is continuous and some regularization is needed to
compute the infinite sums in the Casimir energy (\ref{eq:cas}).
Over the last few years it became clear that the correct
regularization procedure is based on a cut-off in the number of
fluctuation modes, a number that must be taken to be \lq\lq equal"
in the vacuum and kink sectors, see \cite{Rebhan}. More recently,
controversy arose about whether an energy cut-off regularization
can yield the correct answer, see \cite{Flores} and \cite{vanN}.
This is a subtle matter indeed, but all the evidence leads us to
rely on the mode number regularization method.

Before deriving (\ref{eq:dhn}), we outline the set-up pointed out
in \cite{Barton} that underlies our approach below. The problem is
discretised by confining the system to a box of length $2 L$,
which is much greater than any length scale characteristic of the
potential term or of any of its bound states, and the limit
$L\rightarrow \infty$ is taken at the end of the analysis. In
order to specify the boundary conditions we distinguish between
(a) eigenstates with eigenvalue greater than $v^2$, giving rise to
the continuous part of the spectrum, and (b) those with eigenvalue
lower than or equal to $v^2$, corresponding respectively to bound
or half-bound states. Fictitious boundary conditions $\psi(\pm
L)=0$ are imposed on the eigenstates of type (a), but we can
safely anticipate the limit $L\rightarrow\infty$ for states of
type (b) and use for them the eigenfunctions of the actual bound
or half-bound states respectively decrease exponentially or go to
a finite constant at long distances. In any case, we stress the
fact that the effects induced by the boundary conditions disappear
when the $L\rightarrow\infty$ limit is taken. Nevertheless, we
shall approach first the problem by focusing on operators ${\cal
K}$ without half-bound states. This particular case (tackling
operators which involve potentials without reflection, such as
those in the kink and soliton of the well known $\phi^4$ and
Sine-Gordon model) will be dealt with later.

Apart from restricting the spectra of ${\cal V}$ and ${\cal K}$
\[
{\cal V}\psi[k;x]={\bar\omega}^2(k)\psi[k;x]\qquad , \qquad {\cal
K}\psi[q;x]=\omega^2(q)\psi[q;x]
\]
to those eigenfunctions of type (a) satisfying Dirichlet boundary
conditions $\psi(\pm L)=0$, we shall only consider symmetric
potentials $V(x)=V(-x)$ which allows us the choice of  odd,
$\psi(x)=-\psi(-x)$, and,  even, $\psi(x)=\psi(-x)$, eigenstates
complying with the boundary conditions. This breaks the degeneracy
that arises by imposing periodic conditions on eigenfunctions with
non defined-parity (plane waves), and hence skips some problems
arising when the number of bound states is odd.

A brief description of the eigenvalues and eigenfunctions of
${\cal V}$ and  ${\cal K}$ in a finite box follows.

\begin{enumerate}
\item Spectrum of ${\cal V}$

\begin{itemize}

\item Half-bound state.

The constant function,
\[
\psi_0(x)={\rm constant} \qquad ,
\]
is an eigenfunction of ${\cal V}$ with eigenvalue $v^2$ that do
not grow at $x=\pm\infty$.

\item Odd eigenfunctions

If $n^-$ is a natural number strictly positive and the wave
vectors $k_{n^-}$ satisfy
\[
k_{n^-}\cdot L=\pi n^- \hspace{1cm},\hspace{1cm} n^-\in {\Bbb
N}^+\qquad ,
\]
then the odd functions, $\psi_{n^-}(x)=-\psi_{n^-}(-x)$,
\[
\psi_{n^-}(x)=A_{n^-}\cdot {\rm sin}(k_{n^-}\cdot x)
\]
belong to the spectrum of ${\cal V}$ with eigenvalue
${\bar\omega}^2(k_{n^-})=k_{n^-}^2+v^2$. Since
$(k_{n^-+1}-k_{n^-})L=\pi$ the density of odd eigenstates is:
$\bar\rho^-(k)=\frac{L}{\pi}$.

\item Even eigenfunctions

If $n^+$ is a natural number and the wave vectors $k_{n^+}$
satisfy
\[
k_{n^+}\cdot L=\pi(n^++\textstyle\frac{1}{2}) \hspace{1cm} ,
\hspace{1cm} n^+\in {\Bbb N}\qquad ,
\]
then the even functions, $\psi_{n^+}(x)=\psi_{n^+}(-x)$,
\[
\psi_{n^+}(x)=A_{n^+}\cdot {\rm cos}(k_{n^+}\cdot x)
\]
belong to the spectrum of ${\cal V}$ with eigenvalue
${\bar\omega}^2(k_{n^+})=k_{n^+}^2+v^2$. Now,
$\bar\rho^+(k)=\frac{L}{\pi}$.

\end{itemize}

\item Spectrum of ${\cal K}$
\begin{itemize}
\item Bound states:

We assume that there exist $l$ bound states with $\omega^2\leq
v^2$. In particular we denote by $l^+$ the number of symmetric
bound states and by $l^-$ the number of antisymmetric states,
$l=l^+ + l^-$. If $\omega_{l}^2=v^2$, the highest bound state
becomes a half-bound one and contributes with a weight of
$s_{l}=\frac{1}{2}$ to the mass quantum correction.

\item Odd eigenfunctions.

There are odd eigenfunctions which asymptotically are of the form:
\[
\psi_{n^-}(x) \stackrel{x \rightarrow L}{\simeq} A_{n^-}\cdot{\rm
sin}[q_{n^-}\cdot x+\delta^-(q_{n^-})] \qquad .
\]
The odd phase shifts $\delta^-(q_{n^-})$ are defined in terms of
the transmission and reflection coefficients, see Section \S 3,
but now the boundary conditions require:
\begin{equation}
q_{n^-}\cdot L+\delta^-(q_{n^-})=\pi n^- \hspace{1cm},\hspace{1cm}
n^-\in {\Bbb N}\qquad . \label{eq:bouno}
\end{equation}
Unitarity, transparency at $q=\infty$, and, continuity of the wave
function at $q=0$ (threshold) allow us to set the phase ambiguity
as in \cite{Barton}:
\[
\lim_{q\rightarrow\infty}\delta^-(q)=0 \qquad , \qquad
\lim_{q\rightarrow 0^+}\delta^-(q)=\pi\l^- \qquad .
\]

The question arises: what is the minimum value $n^-=n_0^-$? For
very large $L$, $q_{n_0^-}$ is very small and (\ref{eq:bouno})
becomes:
\begin{equation}
q_{n_0^-}\cdot (L+\delta^{\prime -}(0^+))+\delta^-(0^+)=\pi
 n_0^- \qquad .
\end{equation}
Here, $\delta^{\prime -}(0^+)$ is the limit of the left-hand
derivative of the antisymmetric phase shift when $q$ goes to zero
from the right. The minimum value of $n^-=n_0^-$ is therefore
$n_0^-=l^-$. Since
$(q_{n^-+1}-q_{n^-})L+\delta^-(q_{n^-+1})-\delta(q_{n^-})=\pi$ the
density of antisymmetric states is:
\[
\rho^-(q)=\frac{L}{\pi}+\frac{1}{\pi}\frac{\partial\delta^-(k)}{\partial
k}\qquad .
\]

\item Even eigenfunctions.

There are even eigenfunctions which asymptotically are of the
form:
\[
\psi_{n^+}(x)\stackrel{x \rightarrow L}{\simeq} A_{n^+}\cdot{\rm
cos}[q_{n^+}\cdot x+\delta^+(q_{n^+})] \qquad .
\]
The phase shifts $\delta^+(q_{n^+})$ are defined in terms of the
transmission and reflection coefficients, see Section \S 3, and
the boundary conditions require:
\begin{equation}
q_{n^+}\cdot L+\delta^+(q_{n^+})=\pi(n^++\textstyle\frac{1}{2})
\hspace{1cm},\hspace{1cm} n^+\in {\Bbb N}\qquad .
\end{equation}
Again we set the phase ambiguity by invoking Levinson's theorem as
in \cite{Barton}:
\[
\lim_{q\rightarrow\infty}\delta^+(q)=0 \qquad , \qquad
\lim_{q\rightarrow 0^+}\delta^+(q)=\pi(\l^+-\textstyle\frac{1}{2})
\]
where $l^+$ is the number of even bound states.

From
\[
q_{n_0^+}\cdot (L+\delta^{\prime +}(0^+))+\delta^+(0^+)=\pi
(n_0^++\textstyle\frac{1}{2}) \qquad ,
\]
we read the minimum possible value of $q_{n_0^+}$: $n_0^+=l^+$.
Now,
\[
\rho^+(q)=\frac{L}{\pi}+\frac{1}{\pi} \frac{\partial
\delta^+(k)}{\partial k}\qquad .
\]
We stress that due to the parity of the eigenfunctions there is no
need to consider negative values of $n^{\pm}$.
\end{itemize}
\end{enumerate}

The regularized version of formula (\ref{eq:cas}) reads:
\begin{eqnarray*}
\Delta_1\varepsilon^K&=&\frac{\hbar m}{2}\sum_{i=1}^{l}\omega_i+\\
&+&\frac{\hbar m}{2}\left[\sum^{N}_{n^+=l^+}
\sqrt{q_{n^+}^2+v^2}-\sum^{N_0}_{n^+=0}\sqrt{k_{n^+}^2+v^2}\right]
+ \\ &+&\frac{\hbar
m}{2}\left[\sum^{N'}_{n^-=l^-}\sqrt{q_{n^-}^2+v^2}-
\sum^{N_0'}_{n^-=0}\sqrt{k_{n^-}^2+v^2}\right]\qquad .
\end{eqnarray*}
The first row collects the contributions of the $l=l^++l^-$ bound
states of ${\cal K}$. In the other two rows, the contributions of
even and odd eigenfunctions in the continuous spectra are
accounted for. Note that we have added the contribution of the
constant mode of ${\cal V}$ to the odd sector because it
corresponds to the lower eigenvalue of ${\cal V}$. We stick to
mode number cut-off regularization, i.e. we impose $N=N_0$ and
$N'=N_0'$. This choice balances the number of states of ${\cal K}$
and ${\cal V}$ involved in the above formula. We have
distinguished the number of symmetric and antisymmetric states,
$N$ and $N'$, although we shall consider either $N'=N+1$ or
$N'=N$.

To measure the contribution of ${\cal K}$ with respect to the
contribution of ${\cal V}$ to the Casimir energy mode by mode, we
write the second row in the form:
\begin{eqnarray*}
&&\frac{\hbar m}{2}\left[\sum_{n^+=0}^{N}
\left(\sqrt{q_{n^+}^2+v^2}-\sqrt{k_{n^+}^2+v^2}\right)
-\sum_{n^+=0}^{l^+-1}\sqrt{q_{n^+}^2+v^2}\right]\\
&{\simeq}&\frac{\hbar m}{2}\left[-\sum_{n^+=0}^{N}
\frac{k_{n^+}\delta^+(k_{n^+})}{L\sqrt{k_{n^+}^2+v^2}}
-\sum_{n^+=0}^{l^+-1}\sqrt{q_{n^+}^2+v^2}+
\theta(\textstyle\frac{1}{L^2})\right]\\
&\stackrel{{L\rightarrow\infty}
\atop{N\rightarrow\infty}}{\simeq}&-\frac{\hbar
m}{2}\left[\int_0^\infty \,\frac{dk}{\pi}\,
\frac{d\omega(k)}{dk}\delta^+(k)+l^+v \right]\qquad .
\end{eqnarray*}
A similar process for the odd eigenfunctions gives:
\begin{eqnarray*}
&&\frac{\hbar
m}{2}\left[\sum_{n^-=0}^{N'}\left(\sqrt{q_{n^-}^2+v^2}
-\sqrt{k_{n^-}^2+v^2}\right)
-\sum_{n^-=0}^{l^--1}\sqrt{q_{n^-}^2+v^2}\right]\\ &{\simeq} &
\frac{\hbar m}{2}\left[-\sum_{n^-=0}^{N'}
\frac{k_{n^-}\delta^-(k_{n^-})}{L\sqrt{k_{n^-}^2+v^2}}
-\sum_{n^-=0}^{l^--1}\sqrt{q_{n^-}^2+v^2}+\theta(\textstyle\frac{1}{L^2})\right]\\
&\stackrel{{L\rightarrow\infty}\atop
{N'\rightarrow\infty}}{\simeq}& -\frac{\hbar
m}{2}\left[\int_0^\infty \,\frac{dk}{\pi}\,
\frac{d\omega(k)}{dk}\delta^-(k)+l^-v \right]\qquad .
\end{eqnarray*}
The sum of all contributions plus a partial integration,
\begin{eqnarray*}
\Delta_1 \varepsilon^{\rm K}&=& \frac{\hbar m}{2}  \sum_{i=1}^{l}
\omega_i - \\ &&-\frac{\hbar m}{2} l^+ v -\left.\frac{\hbar
m}{2\pi} \delta^+(k)\omega(k)\right|_0^\infty +\frac{\hbar m}{2
\pi}\int_0^\infty dk \frac{d\delta^+(k)}{dk}\omega(k) \\
&&-\frac{\hbar m}{2}l^-v -\left.\frac{\hbar m}{2\pi}
\delta^-(k)\omega(k)\right|_0^\infty +\frac{\hbar
m}{2\pi}\int_0^\infty dk \frac{d\delta^-(k)}{dk}\omega(k)\qquad ,
\end{eqnarray*}
and the asymptotic behaviour of the total phase shift,
\[
\delta^+(k) \stackrel{k\rightarrow\infty}{\cong}
\frac{1}{k}\int_0^\infty dx V(x) \cos^2(k x) \hspace{1cm}
\delta^-(k) \stackrel{k\rightarrow\infty}{\cong}
\frac{1}{k}\int_0^\infty dx V(x) \sin^2(k x)
\]
show that:
\begin{eqnarray*}
\Delta_1 \varepsilon^{\rm K}&=& \frac{\hbar m}{2} \sum_{i=1}^{l}
\omega_i -\frac{\hbar m}{2} l^+\, v +\frac{\hbar m}{2\pi}
\delta^+(0) v-\frac{\hbar m}{2} l^-\, v +\frac{\hbar m}{2\pi}
\delta^-(0) v -\\ && -\frac{\hbar m}{2\pi} \int_0^\infty dx V(x)
[\cos^2(kx)+\sin^2(kx)] +\frac{\hbar m}{2\pi}\int_0^\infty dk
\frac{d [\delta^+(k)+\delta^-(k)]}{dk}\omega(k) \qquad .
\end{eqnarray*}
From the 1D Levinson theorem we notice that the two last terms in
the first line of the above formula cancel each other but the two
preceding ones leave a contribution: $-\frac{v}{2}$. Therefore,
the formula for the kink Casimir energy reads:
\begin{equation}
\Delta_1\varepsilon^{\rm K}=\frac{\hbar
m}{2}\left[\sum_{i=1}^{l}\omega_i-\frac{v}{2}+\frac{1}{\pi}\int_0^\infty
\!\! dk
\frac{d\delta(k)}{dk}\omega(k)-\frac{\left<V(x)\right>}{2\pi}\right]\qquad
, \label{onel}
\end{equation}
where $\left<V(x)\right>=\int_{-\infty}^\infty V(x) dx =2
\int_0^\infty V(x) dx$ is the expectation value of the potential
term and the contribution of the continuous spectrum is encoded in
the total phase shift: $\delta(k)=\delta^+(k)+\delta^-(k)$. We
call attention to the contribution $-\frac{\hbar mv}{4}$, which
tells us that the contribution of the mode with $k=0$ wave vector
in ${\rm Spec}{\cal V}$ is half the contribution of a bona fide
bound state -hence, the name-.

The generalization of this approach to tackle the presence of a
bound state with $k=0$ and $\omega^2=v^2$ in ${\rm Spec}{\cal K}$
(half-bound state) is direct. The Levinson theorem,
\begin{equation}
\delta_-(0^+)=l_-\pi \qquad , \qquad
\delta_+(0^+)=l_+\pi-\frac{\pi}{2} \qquad , \label{eq:Levi}
\end{equation}
works as above, but there is now a half-bound state that
contributes with $\frac{1}{2}$ to both the number $l=l_-+l_+$ and
the mass quantum correction $\Delta_1 \varepsilon^K$, see
\cite{Graham}. Thus, we write (\ref{eq:Levi}) as $\delta(0^+)=l
\pi-\frac{\pi}{2}=n_b\pi$, and, bearing in mind the previous
observation, the above calculations reproduce the results obtained
in Reference \cite{Rebhan}. The tricky point to realize is how the
contribution of the threshold state turns (\ref{eq:Levi}) into
$\delta(q^+)=n_b \pi$, with $n_b$ the number of bound states with
$\omega^2 < v^2$. In the kink mass formula, the contribution of
the half-state would cancel out the term $-\frac{v}{2}$, and
everything is okay. For this reason, the formula shown in
\cite{Rebhan} is valid only for operators involving potentials
without reflection. A final expression involving both of these
cases provides the first row of formula (\ref{eq:dhn})
\begin{equation}
\Delta_1\varepsilon^{\rm K}=\frac{\hbar
m}{2}\left[\sum_{i=1}^{l-1}\omega_i+s_l \omega_l
-\frac{v}{2}+\frac{1}{\pi}\int_0^\infty \!\! dk
\frac{d\delta(k)}{dk}\omega(k)-\frac{\left<V(x)\right>}{2\pi}\right]\qquad
, \label{one2}
\end{equation}
where $s_l=\frac{1}{2}$ if we deal with a reflectionless potential
but $s_l=1$ if the reflection coefficient is non-zero.

\section*{Acknowledgements}

E-mail correspondence with A. Rebhan, P. van Niewenhuizen and R.
Wimmer about important points of their work on kink quantization
is gratefully acknowledged. We also thank J.J. Blanco-Pillado for
critical reading of the manuscript.

A.A.I. thanks to the  Secretaria de Estado de Educaci\'on y
Universidades of Spain for financial support. The work of W.G.F.
has been partially supported by Oviedo University, Vicerrectorado
de Investigacion grantMB-03-514-1.

\end{document}